\documentclass[12pt,onecolumn,draftcls]{IEEEtran}
\usepackage{bbm}
\usepackage{dsfont}
\usepackage{yfonts}
\usepackage{bm,cleveref}
\usepackage{amsmath}
\usepackage{amssymb}
\usepackage{amsfonts}
\usepackage{mathrsfs} 
\usepackage{stfloats}
\usepackage{amsthm}
\usepackage{subfigure}
\usepackage{graphicx}
\usepackage{epstopdf,amssymb,amsmath}
\usepackage{multicol}
\usepackage[noadjust]{cite}
\usepackage{setspace}
\usepackage{stfloats,subfigure}
\usepackage{midfloat}
\usepackage[normal]{threeparttable}
\usepackage{amsthm}
\usepackage{cuted}
\usepackage{cases,subeqnarray}
\usepackage{bm,multirow,bigstrut}
\usepackage{textcomp}
\usepackage{latexsym,bm}
\usepackage{booktabs,changebar}
\usepackage{xcolor}
\usepackage{mathtools}
\usepackage{dsfont}
\usepackage{extarrows}
\ifCLASSOPTIONcompsoc
\usepackage[nocompress]{cite}
\else
\usepackage{cite}
\fi

\begin{document}

\title{Performance Analysis of Mixed-ADC Massive MIMO Systems over Rician Fading Channels}

\author{Jiayi~Zhang,~\IEEEmembership{Member,~IEEE,}
        Linglong~Dai,~\IEEEmembership{Senior Member,~IEEE},
        Ziyan~He,\\
        Shi Jin,~\IEEEmembership{Member,~IEEE},
        and~Xu~Li

\thanks{J. Zhang and X. Li are with the School of Electronic and Information Engineering, Beijing Jiaotong University, Beijing 100044, P. R. China (e-mail: jiayizhang@bjtu.edu.cn).}
\thanks{L. Dai and Z. He are with Department of Electronic Engineering, Tsinghua University, Beijing 100084, P. R. China. (e-mail: daill@tsinghua.edu.cn)}
\thanks{S. Jin is with the National Mobile Communications Research Laboratory, Southeast University, Nanjing 210096, P. R. China (e-mail: jinshi@seu.edu.cn).}
\thanks{Simulation codes are provided to reproduce the results presented in this paper: http://oa.ee.tsinghua.edu.cn/dailinglong/.}
}

\maketitle
\vspace{-5mm}
\begin{abstract}
The practical deployment of massive multiple-input multiple-output (MIMO) in future fifth generation (5G) wireless communication systems is challenging due to its high hardware cost and power consumption. One promising solution to address this challenge is to adopt the low-resolution analog-to-digital converter (ADC) architecture. However, the practical implementation of such architecture is challenging due to the required complex signal processing to compensate the coarse quantization caused by low-resolution ADCs. Therefore, few high-resolution ADCs are reserved in the recently proposed mixed-ADC architecture to enable low-complexity transceiver algorithms. In contrast to previous works over Rayleigh fading channels, we investigate the performance of mixed-ADC massive MIMO systems over the Rician fading channel, which is more general for the 5G scenarios like Internet of Things (IoT). Specially, novel closed-form approximate expressions for the uplink achievable rate are derived for both cases of perfect and imperfect channel state information (CSI). With the increasing Rician $K$-factor, the derived results show that the achievable rate will converge to a fixed value. We also obtain the power-scaling law that the transmit power of each user can be scaled down proportionally to the inverse of the number of base station (BS) antennas for both perfect and imperfect CSI. Moreover, we reveal the trade-off between the achievable rate and energy efficiency with respect to key system parameters including the quantization bits, number of BS antennas, Rician $K$-factor, user transmit power, and CSI quality. Finally, numerical results are provided to show that the mixed-ADC architecture can achieve a better energy-rate trade-off compared with the ideal infinite-resolution and low-resolution ADC architectures.
\end{abstract}

\begin{IEEEkeywords}
Achievable rate, mixed-ADC receiver, massive MIMO, Rician fading channels.
\end{IEEEkeywords}

\IEEEpeerreviewmaketitle

\section{Introduction}
The research on the fifth generation (5G) wireless communication systems has been undertaken in recent years to meet the demanding requirement of around 1000 times data traffic (e.g., see \cite{agiwal2016next,Andrews2014what,Boccardi2014five} and references therein). Massive multiple-input multiple-output (MIMO) has emerged as one of the most promising technologies to further improve the achievable rate and energy efficiency by ten times. By using hundreds or even thousands of antennas at the base station (BS), the inter-user interference in massive MIMO systems can be substantially reduced by simple signal processing such as maximum-ratio combining (MRC) and zero-forcing (ZF) schemes.

For practical implementation of massive MIMO, a large number of antennas significantly complicate the hardware design. It is difficult to use a dedicated radio frequency (RF) chain for every antenna, since the associated hardware cost and power consumption is unaffordable when the number of BS antennas becomes very large. Several massive MIMO architectures have been proposed to address this challenging problem. The first one is the hybrid analog and digital precoding scheme, which uses different analog approaches like phase shifters \cite{gao2016energy}, switches \cite{mendez2016hybrid} or lenses \cite{brady2013beamspace} to substantially reduce the number of RF chains. However, the performance of such hybrid precoding system highly depends on the accurate control of the analog components. Moreover, if the beamwidth is small, the fast acquisition and alignment of the optimal beam in millimeter wave channels is challenging \cite{Heath2016overview}.

On a parallel avenue, the high-resolution ADC (e.g., more than 10-12 bits for commercial use) in the RF chain is a power-hungry device, especially when large signal bandwidth is considered. This is due to the factor that the power consumption of a typical ADC roughly scales linearly with the signal bandwidth and exponentially with the quantization bit. Moreover, a large demand on digital signal processing with high-resolution ADCs will induce excessive circuitry power consumption and hardware cost. Therefore, an alternative massive MIMO architecture at the receiver is to reduce the high resolution (e.g., 10-12 bits) of the ADCs to low resolution (e.g., 1-3 bits) while keeping the number of RF chains unchanged. However, the severe nonlinear distortion caused by low-resolution ADCs inevitably causes several problems, including capacity loss at high SNR \cite{mo2015capacity,zhang2016spectral}, high pilot overhead for channel estimation \cite{mo2016channel,Studer2016quantized}, error floor for multi-user detection \cite{wen2016bayes,choi2016near,palguna2016millimeter}, and complex precoder design \cite{gokceoglu2016spatio,Roth2016achievable,stein2015overdemodulation}.

\subsection{Related Works}
In order to alleviate aforementioned concerns, a mixed-ADC architecture for massive MIMO systems has been recently proposed in \cite{liang2016mixed}, where a small number of high-resolution ADCs are employed, while the other RF chains are composed of one-bit ADCs. This architecture has the potential of designing practical massive MIMO systems while still maintaining a large fraction of the performance gain. For example, the reserved high-resolution ADCs can be utilized to realize accurate channel estimation in a round-robin manner with reduced-complexity algorithm and affordable pilot overhead \cite{liang2016mixed}. Furthermore, the deduced channel estimation error is Gaussian distributed, which facilitates the performance analysis and front-end designs of massive MIMO systems.

While there have been considerable works on homogeneous-ADC architectures with low ADC resolutions, very little attention has been paid to the promising mixed-ADC massive MIMO system. Only the analysis of generalized mutual information has been emphasized in \cite{liang2016mixed} to demonstrate that the mixed-ADC architecture is able to achieve a large fraction of the channel capacity of ideal ADC architectures over frequency-flat channels. This analysis is then extended to frequency-selective fading channels \cite{liang2016frequency}. In \cite{zhang2016mixed}, a generic Bayes detector is proposed to achieve the optimal signal detection in the mean squared error (MSE) sense by using the generalized approximate message passing algorithm. The authors in \cite{tan2016spectral} derived a closed-form approximation of the achievable rate of mixed-ADC massive MIMO systems with the MRC detector. These pioneer works validate the merits of the mixed-ADC architecture for massive MIMO systems.

The common assumption of the aforementioned works is that massive MIMO systems are operating over Rayleigh fading channels. However, Rician fading, which includes the scenario of line-of-sight (LoS) signal propagation, is more general and realistic than Rayleigh fading. {Moreover, recent works reveal that LoS (or near-LoS) signal propagation is mostly relied on in massive MIMO \cite{zhang2014power,miridakis2016distributed,zhang2016achievable} or millimeter-wave communications \cite{brady2013beamspace,Rangan2014millimeter,liu2016line}.} Unfortunately, a Rician model makes the analysis of massive MIMO systems more complicated, as the extension of current results in Rayleigh fading channels \cite{tan2016spectral} to those in Rician fading channels is not easy. This is mainly because of the inherent difficulty in deriving the simple and mathematically tractable expressions for Rician fading channels. Moreover, channel estimation in massive MIMO is a challenging problem. An erroneous channel estimation may occur due to large number of estimation parameters, rapid time-varying channel fading, and imperfect feedback signaling. Therefore, the issue of imperfect channel state information (CSI) should also be taken into account in the performance analysis of mixed-ADC massive MIMO systems operating over Rician fading channels.

\subsection{Contributions}
This study aims to examine the mixed-ADC massive MIMO system in terms of achievable rate and energy efficiency. Different from most of the existing works in the literature, we model the mixed-ADC massive MIMO links as more general Rician fading channels for performance analysis. We demonstrate that the mixed-ADC architecture is practically useful because it can provide comparable performance with the ideal infinite-resolution ADC architecture, while the signal processing complexity and power consumption can be reduced at the same time. More specifically, the contributions of this work are summarized as follows:

\begin{itemize}
\item With the assumption of Rician fading channels, novel closed-form approximate expressions for the achievable rate of mixed-ADC massive MIMO systems are derived. These results reveal the effects of the number of BS antennas, user transmit power, quantization bits, number of high-resolution ADCs, and Rician $K$-factor on the achievable rate performance. Note that our derived results are generalized, and can include many previously published works \cite{tan2016spectral,zhang2014power,ngo2013energy,fan2015uplink,zhang2016spectral} as special cases by adjusting system parameters accordingly.
\item The power-scaling law of the mixed-ADC architecture under Rician fading channels is elaborated for both perfect and imperfect CSI cases. Our results show that when the number of BS antennas, $M$, gets asymptotically large, the transmit power of each user can be scaled down by $1/M$ to obtain a desirable rate for both perfect and imperfect CSI cases, if the Rician $K$-factor is nonzero. This new finding is different from the result in Rayleigh fading channels \cite{ngo2013energy,tan2016spectral,fan2015uplink}. {In contrast to \cite{zhang2016spectral}, we focus on the mixed-ADC architecture and prove that the derived rate limits depend on the proportion of the high-resolution ADCs $\kappa$ and the total number of low- and high-resolution ADCs.}
\item By adopting a generic power consumption model for the mixed-ADC architecture, we quantify the trade-off between the achievable rate and energy efficiency for different number of high-resolution ADCs and quantization bits. The optimal number of quantization bits and BS antennas needed to maximize the energy efficiency are illustrated. We further prove that the mixed-ADC architecture can work better under stronger Rician fading channels and larger user transmit power.
\end{itemize}

\subsection{Outline}
The rest of this paper is organized as follows. The system model is briefly introduced in Section \ref{se:system}. Then, the close-form approximate expressions for uplink achievable rate of mixed-ADC architectures are derived in Section \ref{se:achievable_rate}. The power-scaling laws for both perfect and imperfect CSI cases are also derived in Section \ref{se:achievable_rate}. {In Section \ref{se:numerical_results}, we study the energy efficiency under a practical power consumption model. Moreover, numerical evaluations are provided in Section \ref{se:numerical_results} to illustrate the achievable rate and energy efficiency performance of the mixed-ADC architecture with respect to various system parameters.} Finally, conclusions are drawn in Section \ref{se:conclusion}.

\subsection{Notation}
In this paper, we use $\boldsymbol{\rm x}$ and $\boldsymbol{\rm X}$ in bold typeface to represent vectors and matrices, respectively, while scalars are presented in normal typeface, such as $x$. We use $\boldsymbol{\rm X}^T$ and $\boldsymbol{\rm X}^H$ to represent the transpose and conjugate transpose of a matrix $\boldsymbol{\rm X}$, respectively. $\boldsymbol{\rm I}_N$ stands for an $N \times N$ identity matrix, and $\|\boldsymbol{\rm X}\|_F$ denotes the Frobenius norm of a matrix $\boldsymbol{\rm X}$. Moreover, $\mathbb{E}\{\cdot\}$ is the expectation operator, and $\boldsymbol{\rm x} \sim \mathcal{CN}(\boldsymbol{\rm m},\sigma^2 \boldsymbol{\rm I} )$ denotes that $\boldsymbol{\rm x}$ is a circularly symmetric complex Gaussian stochastic vector with mean matrix $\boldsymbol{\rm m}$ and covariance matrix $\sigma^2 \boldsymbol{\rm I}$.

\section{System Model}\label{se:system}
\begin{figure}[!t]
\centering
\includegraphics[scale=1]{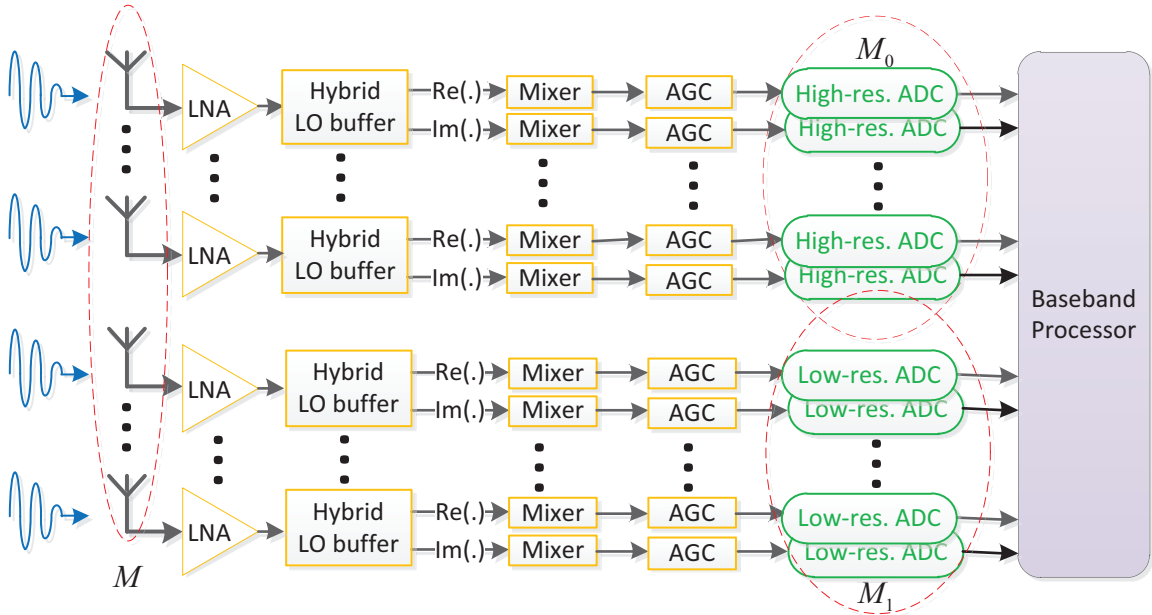}
\caption{System model of mixed-ADC massive MIMO receivers with ${M_0}$ pairs of high-resolution ADCs and ${M_1}$ pairs of low-resolution ADCs. The number of receive antennas is $M$, and $M={M_0}+{M_1}$.}
\label{mixed_ADC}
\end{figure}

In this paper, we consider a uplink single-cell massive MIMO system with $M$ antennas at the BS and $N\left( {N \ll M} \right)$ single-antenna users \cite{agiwal2016next,Andrews2014what,Boccardi2014five}. Since the hardware cost and energy consumption scale linearly with the number of BS antennas, a mixed-ADC architecture as illustrated in Fig. \ref{mixed_ADC} is employed at the BS instead of costly high-resolution ADCs. Note that the transmit antenna of each user is connected with high-resolution digital-to-analog converters (DACs). In the mixed-ADC architecture, ${M_0}$ pairs of high-resolution ADCs (e.g., 10-12 bits) are connected to ${M_0}$ BS antennas, while all other ${M_1}$ pairs of low-resolution ADCs (e.g., 1-3 bits) are connected to the remained ${M_1}$ BS antennas. Note that ${M_1} = M - {M_0}$, and we define $\kappa  \buildrel \Delta \over = {{{M_0}} \mathord{\left/{\vphantom {{{M_0}} M}} \right.\kern-\nulldelimiterspace} M}$ $\left( {0 \le \kappa  \le 1} \right)$ as the proportion of the high-resolution ADCs in the mixed-ADC architecture.

As illustrated in Fig. \ref{mixed_ADC}, the power consumption of the receiver is induced by the RF chain and the baseband processor. The local oscilator (LO), which is used with a mixer to change the frequency of a signal, is shared by all RF chains. Moreover, each RF chain is composed of a low-noise amplifier (LNA), a $\pi/2$ hybrid and the LO buffer. The in-phase and quadrature circuit consist of a mixer including a quadrature-hybrid coupler, an ADC, and an automatic gain control (AGC), respectively, Note that the AGC is used to amplify the received signal to use the full dynamic range of the ADC. The use of one-bit ADCs can simplify the RF chain, e.g., an AGC is not necessary because only the sign of the input signal is output.

Let us consider the channel between each transmit-receive antenna pair as a block-fading channel, i.e., a channel that stays unchanged for several channel uses, and changes independently from block to block \cite{ngo2013energy}. In addition, $\boldsymbol{\rm G}$ is the $M \times N$ channel matrix between the $M$-antenna BS and the $N$ single-antenna users, which can be modeled as \cite{zhang2014power}
\begin{equation}
\boldsymbol{\rm G} = \boldsymbol{\rm H}{{\boldsymbol{\rm D}}^{{1 \mathord{\left/
 {\vphantom {1 2}} \right.
 \kern-\nulldelimiterspace} 2}}},
\end{equation}
where the diagonal matrix $\boldsymbol{\rm D} \in {\mathbb{C}^{N \times N}}$ denotes the large-scale channel fading (geometric attenuation and log-normal shadow fading) with the diagonal elements ${\beta _n}$'s given by ${\left[ \boldsymbol{\rm D} \right]_{nn}} = {\beta _n}, (n=1,\cdots, N)$ and $\boldsymbol{\rm H} \in {\mathbb{C}^{M \times N}}$ presents the small-scale fading channel matrix. In contrast to \cite{tan2016spectral}, we consider Rician fading channels, which includes a deterministic component $\boldsymbol{\rm \bar H}$ corresponding to the LoS signal and a Rayleigh-distribution random component ${\boldsymbol{\rm H}_\omega }$ accounting for the scattered signals as \cite{zhang2016spectral}
\begin{equation}
\boldsymbol{ \rm H} = \boldsymbol{\bar {\rm H}}\sqrt {\boldsymbol{\Omega} {{\left( {\boldsymbol{\Omega}  + {{\boldsymbol{\rm I}}_N}} \right)}^{ - 1}}}  + {\boldsymbol{\rm H}_\omega }\sqrt {{{\left( {\boldsymbol{\Omega}  + {{\boldsymbol{\rm I}}_N}} \right)}^{ - 1}}} ,
\end{equation}
where ${\boldsymbol{\rm H}_\omega } \sim \mathcal{CN}\left( {\boldsymbol{0},{{\boldsymbol{\rm I}}_N}} \right)$, and $\boldsymbol{\bar {\rm H}}$ with an arbitrary rank can be expressed as \cite{zhang2014power}
\begin{equation}
{\left[ {\boldsymbol{\bar {\rm H}}} \right]_{mn}} = \exp \left( { - j\left( {m - 1} \right)kd\sin {\theta _n}} \right),
\end{equation}
where $\theta _n$ denotes the arrival of angle (AoA) of the $n$th user, $k = {{2\pi d} \mathord{\left/
 {\vphantom {{2\pi d} \lambda }} \right.
 \kern-\nulldelimiterspace} \lambda }$
with $\lambda $ being the wavelength and $d$ being the inter-antenna space. In addition, the elements of the diagonal matrix $\bf{\Omega} $ is the Rician $K$-factor ${K_n}$, which denotes the ratio of the power of the deterministic component to the power of the scattered components.

We define $\boldsymbol{{\rm G}}_0$ as the ${M_0} \times N$ channel matrix from the $N$ users to the ${M_0}$ BS antennas connected with high-resolution ADCs, and $\boldsymbol{{\rm G}}_1$ as the ${M_1} \times N$ channel matrix from the $N$ users to the ${M_1}$ BS antennas connected with low-resolution ADCs. Therefore, we have
\begin{equation}
\boldsymbol{\rm G} = \left[ {\begin{array}{*{20}{c}}
  {{\boldsymbol{\rm G}_0}} \\
  {{\boldsymbol{\rm G}_1}}
\end{array}} \right].
\end{equation}
Let ${p_u}$ be the transmit power of each user and $\boldsymbol {\rm x} \in {\mathbb{C}^{N \times 1}}$ be the transmit vector for all $N$ users. The received signals at the $2{M_0}$ high-resolution ADCs can be written as
\begin{equation}\label{test1}
\boldsymbol {\rm y_0} = \sqrt {{p_u}}\boldsymbol  {\rm G}_0\boldsymbol  {\rm x} + \boldsymbol  {\rm n}_0,
\end{equation}
where ${p_u}$ is the transmit power of each user, $\boldsymbol {\rm x} \in {\mathbb{C}^{N \times 1}}$ is the transmitted signal vector for all $N$ users, and $ \boldsymbol{\rm n}_0\sim \mathcal{CN}\left( {\boldsymbol{0},{{\boldsymbol{\rm I}}_{M_0}}} \right)$ denotes the additive white Gaussian noise (AWGN) matrix with independent and identically distributed (i.i.d.) components following the distribution $\mathcal{CN}(0,1)$. We further follow the general assumption that the transmitted signal vector $\boldsymbol {\rm x}$ is Gaussian distributed, the quantization noise is the worst case of Gaussian distributed, and the quantization error of low-resolution ADCs can be well approximated as a linear gain with the additive quantization noise model (AQNM) \cite{bai2015energy}. Note that the AQNM model is accurate enough at low and medium SNRs, which has been widely used in quantized MIMO systems \cite{bai2015energy,fletcher2007robust,mo2016hybrid,zhang2016spectral,fan2015uplink}. Thus, the quantized received signals at the output of $2{M_1}$ low-resolution ADCs can be expressed as
\begin{align}\label{test2}
\boldsymbol {\rm y}_1 &= \mathbb{Q}\left( {{\boldsymbol {\rm {\tilde  y}}_1}} \right) \approx \alpha {\boldsymbol {\rm {\tilde y}}_1} +\boldsymbol {\rm  n_q} \notag \\
&= \sqrt {{p_u}} \alpha {\boldsymbol {\rm G}_1}\boldsymbol {\rm x }+ \alpha {\boldsymbol{\rm n}_1} + \boldsymbol {\rm  n_q},
\end{align}
where ${{\boldsymbol {\rm {\tilde  y}}_1}} = \sqrt {{p_u}} {\boldsymbol{\rm G}_1} + {\boldsymbol{\rm n}_1}$ denotes the uplink received signal vector at the $2{M_1}$ low-resolution ADCs, $\mathbb{Q}\left( \cdot \right)$ is the scalar quantization function which applies component-wise and separately to the real and imaginary parts, $\boldsymbol {\rm  n_q}$ denotes the additive Gaussian quantization noise vector which is uncorrelated with $\boldsymbol {\rm y}_1$, and $\alpha  = 1 - \rho $ is a linear gain with \cite[Eq. (13)]{fletcher2007robust}
\begin{align}\label{equation1}
{\rho  = \frac{{\mathbb{E}\left\{ {{{\| {{\boldsymbol {\rm {\tilde  y}}_1} - {\boldsymbol {\rm y}_1}} \|}^2}} \right\}} } { {\mathbb{E}\left\{ {{{\| {{\boldsymbol {\rm {\tilde  y}}_1}} \|}^2}} \right\}}}}
\end{align}
denoting the distortion factor of the low-resolution ADC. The exact values of $\rho$ have been given in Table I with respect to different resolution $b$'s \cite{max1960quantizing}. For large quantization bits (e.g., $b>5$), the distortion factor $\rho$ can be approximated as \cite{max1960quantizing}
\begin{equation}\label{appro_distortion_factor}
\rho  \approx \frac{\pi \sqrt{3}}{2} 2^{-2b}.
\end{equation}
From (\ref{test2}) and (\ref{equation1}), the covariance matrix of $\boldsymbol {\rm  n_q}$ is given by
\begin{equation}\label{eq:covariance}
{\boldsymbol {\rm R}_{{\boldsymbol {\rm  n_q}}}} = \alpha \rho {\rm{diag}}\left( {{p_u}{\boldsymbol {\rm G}_1}{\boldsymbol {\rm G}_1^H} + {\boldsymbol {{\rm I}}_{{M_1}}}} \right).
\end{equation}
By considering (\ref{test1}) and (\ref{test2}), the overall received signal at the BS can be expressed as
\begin{equation} \label{test3}
{ {\boldsymbol {\rm y} = }}\left[ {\begin{array}{*{20}{c}}
{{{\boldsymbol {\rm{y}}}_0}}\\
{{\boldsymbol {\rm y}_1}}
\end{array}} \right] \approx \left[ {\begin{array}{*{20}{c}}
{\sqrt {{p_u}}\boldsymbol  {\rm G}_0\boldsymbol  {\rm x} + \boldsymbol  {\rm n}_0}\\
{ \sqrt {{p_u}} \alpha {\boldsymbol {\rm G}_1}\boldsymbol {\rm x }+ \alpha {\boldsymbol{\rm n}_1} + \boldsymbol {\rm  n_q}}
\end{array}} \right].
\end{equation}

\begin{table}[!t]
\renewcommand{\thetable}{\Roman{table}}
\caption{Distortion Factors for Different Quantization Bits \cite{max1960quantizing}}
\label{table}
\centering
\begin{tabular}{cccccc}
\toprule[2pt]
  $b$ & 1 & 2     & 3  & 4    & 5 \\
\midrule[1pt]
  $\rho$     &  0.3634      &   0.1175 &   0.03454 & 0.009497 & 0.002499  \\
 \bottomrule[2pt]
\end{tabular}
\end{table}

\section{Achievable Rate}\label{se:achievable_rate}
In this section, we derive approximate closed-form expressions for the uplink achievable rate of mixed-ADC massive MIMO systems. Both cases of perfect and imperfect CSI at the BS receiver are considered. The MRC receiver is assumed to be employed at the BS, as such receiver is simple and near optimal when the number of BS antennas becomes large \cite{ngo2013energy}.

\subsection{Perfect CSI}
We first consider the case of perfect CSI at the BS. By using (\ref{test3}), the detected signal vector after MRC is given by
\[{ {\boldsymbol {\rm r} = }}{{\boldsymbol  {\rm G}}^H}{ {\boldsymbol {\rm y} = }}{\left[ {\begin{array}{*{20}{c}}
{{{\boldsymbol  {\rm G}}_0}}\\
{{{\boldsymbol  {\rm G}}_1}}
\end{array}} \right]^H}\left[ {\begin{array}{*{20}{c}}
{\sqrt {{p_u}}\boldsymbol  {\rm G}_0\boldsymbol  {\rm x} + \boldsymbol  {\rm n}_0}\\
{ \sqrt {{p_u}} \alpha {\boldsymbol {\rm G}_1}\boldsymbol {\rm x }+ \alpha {\boldsymbol{\rm n}_1} + \boldsymbol {\rm  n_q}}
\end{array}} \right],\]
which can be further expressed as
\begin{align}\label{test4}
{ {\boldsymbol {\rm r} = }}&\sqrt {{p_u}} \left( {{\boldsymbol  {\rm G}_0^H}{{\boldsymbol  {\rm G}}_0} + \alpha {{\boldsymbol  {\rm G}}_1^H}{{\boldsymbol  {\rm G}}_1}} \right){\boldsymbol {\rm{x}}}\! +\! \left( {{{\boldsymbol  {\rm G}_0^H}}{{\boldsymbol  {\rm n}}_0} \!+\! \alpha {{\boldsymbol  {\rm G}}_1^H}{{\boldsymbol  {\rm n}}_1}} \right){\rm{ + }}{{\boldsymbol  {\rm G}}_1^H}{{\boldsymbol {\rm  n_q}}}.
\end{align}
Let $\boldsymbol{\rm{g}}_{n0}$ and $\boldsymbol{\rm{g}}_{n1}$ be the $n$th column of ${\boldsymbol  {\rm G}}_0$ and ${\boldsymbol  {\rm G}}_1$, respectively. Then, the $n$th element of $\boldsymbol {\rm r}$ in \eqref{test4} can be written as
\begin{align}\label{test5}
{r_n} =& \underbrace {\sqrt {{p_u}} \left( {\boldsymbol{\rm{g}}_{n0}^H{\boldsymbol {\rm{g}}_{n0}} + \alpha \boldsymbol{\rm{g}}_{n1}^H{\boldsymbol{\rm{g}}_{n1}}} \right){x_n}}_{\text{desired signal}}  \notag\\
&+ \underbrace {\sqrt {{p_u}} \sum\limits_{ i \ne n}^N {\left( {\boldsymbol{\rm{g}}_{n0}^H{\boldsymbol{\rm{g}}_{i0}} + \alpha \boldsymbol{\rm{g}}_{n1}^H{\boldsymbol{\rm{g}}_{i1}}} \right){x_i}}}_{\text{interuser interference}}  \notag\\
&+ \underbrace {\left( {\boldsymbol{\rm{g}}_{n0}^H{\boldsymbol{\rm{n}}_0} + \alpha \boldsymbol{\rm{g}}_{n1}^H{\boldsymbol{\rm{n}}_1}} \right)}_{\text{AWGN}} + \underbrace {\boldsymbol{\rm{g}}_{n1}^H{{\boldsymbol {\rm  n_q}}}}_{\text{quantization noise}},
\end{align}
where $x_{n}$ is the $n$th element of the signal vector ${\boldsymbol  {\rm x}}$. For a fixed channel realization ${\boldsymbol  {\rm G}}$, the last three terms in \eqref{test5} can be seen as an interference-plus-noise random variable. By assuming the interference-plus-noise follows Gaussian distribution and is independent of ${\boldsymbol  {\rm x}}$ \cite{ngo2013energy,zhang2014power}, we can obtain a lower bound of the achievable rate of the $n$th user as
\begin{equation}\label{test6}
R_{{\rm{P,n}}}^{{\rm{MRC}}} = \mathbb{E}\left\{ {{{\log }_2}\left( {1 + \frac{{{p_u}{{\left| {\boldsymbol{\rm g}_{n0}^H{\boldsymbol{\rm g}_{n0}} + \alpha \boldsymbol{ \rm g}_{n1}^H{\boldsymbol{\rm g}_{n1}}} \right|}^2}}}{{{\Psi _1}}}} \right)} \right\},
\end{equation}
where
\begin{align}\label{eq:inter_noise}
{\Psi _1} &={p_u}\sum\limits_{ i \ne n}^N {{{\left| {{\boldsymbol {\rm{g}}}_{n0}^H{\boldsymbol{\rm g}_{i0}} + \alpha {\boldsymbol{\rm{g}}}_{n1}^H{\boldsymbol{\rm g}_{i1}}} \right|}^2}} + {\left| {{\boldsymbol{\rm{g}}}_{n0}^H{{\boldsymbol{\rm{n}}}_0} + \alpha {\boldsymbol{\rm{g}}}_{n1}^H{{\boldsymbol{\rm{n}}}_1}} \right|^2} \notag \\
 &+ \alpha \rho \boldsymbol{ \rm g}_{n1}^H{\rm{diag}}\left( {{p_u}{\boldsymbol {\rm G}_1}{\boldsymbol {\rm G}_1^H} + {\boldsymbol {{\rm I}}_{{M_1}}}} \right){\boldsymbol{\rm g}_{n1}} \notag \\
&={p_u}\sum\limits_{ i \ne n}^N {{{\left| {{\boldsymbol {\rm{g}}}_{n0}^H{\boldsymbol{\rm g}_{i0}} + \alpha {\boldsymbol{\rm{g}}}_{n1}^H{\boldsymbol{\rm g}_{i1}}} \right|}^2}} + {\left| {{\boldsymbol{\rm{g}}}_{n0}^H{{\boldsymbol{\rm{n}}}_0}}\right|^2} + \alpha{\left|{ {\boldsymbol{\rm{g}}}_{n1}^H{{\boldsymbol{\rm{n}}}_1}} \right|^2} \notag \\
 &+ \alpha \rho {p_u}\boldsymbol{ \rm g}_{n1}^H{\rm{diag}}\left( {{\boldsymbol {\rm G}_1}{\boldsymbol {\rm G}_1^H} } \right){\boldsymbol{\rm g}_{n1}},
\end{align}
where the covariance matrix of $\boldsymbol {\rm  n_q}$ in \eqref{eq:covariance} has been used.
%

For the case of perfect CSI at the BS, the approximate achievable rate is derived in following Lemma 1.
\newtheorem{lemma}{Lemma}
\begin{lemma}  \label{lemma1}
For mixed-ADC massive MIMO systems using MRC receiver with perfect CSI over Rician fading channels, the approximate achievable rate of the $n$th user is given by
\begin{align}\label{test7}
&R_{{\rm{P,n}}}^{{\rm{MRC}}} \approx {\log _2}\Bigg( 1 \notag \\
& \!+\! \frac{{{p_u}{\beta _n}\left[ {\left( {2{K_n} \!+\! 1} \right)\left( {{M_0} + {\alpha ^2}{M_1}} \right) \!+\! {{\left( {{K_n} \!+\! 1} \right)}^2}{{\left( {{M_0} \!+\! \alpha {M_1}} \right)}^2}} \right]}}{{{p_u}\left( {{K_n} \!+\! 1} \right)\sum\limits_{ i \ne n}^N {{\beta _i}\left( {{\Delta _0} \!+\! {\alpha ^2} {\Delta _1}} \right) \!+\! {{\left( {{K_n} \!+\! 1} \right)}^2}\left( {{M_0} \!+\! \alpha {M_1}} \right) \!+\! \alpha \rho {p_u}{\Delta _2}} }} \Bigg),
\end{align}
where
\setcounter{equation}{15}
\begin{equation}
{\Delta _j} \buildrel \Delta \over = \frac{{{K_n}{K_i}\phi _{ni,j}^2 + {M_j}\left( {{K_n} + {K_i} + 1} \right)}}{{{K_i} + 1}} \;\;\; {j = 0,1},
\end{equation}
\begin{equation}
{\phi _{ni,j}} \buildrel \Delta \over = \frac{{\sin \left( {{{{M_j}kd\left( {\sin {\theta _n} - \sin {\theta _i}} \right)} \mathord{\left/
 {\vphantom {{{M_j}kd\left( {\sin {\theta _n} - \sin {\theta _i}} \right)} 2}} \right.
 \kern-\nulldelimiterspace} 2}} \right)}}{{\sin \left( {{{kd\left( {\sin {\theta _n} - \sin {\theta _i}} \right)} \mathord{\left/
 {\vphantom {{kd\left( {\sin {\theta _n} - \sin {\theta _i}} \right)} 2}} \right.
 \kern-\nulldelimiterspace} 2}} \right)}}\;\;\; {j = 0,1},
\end{equation}
\begin{align}
{\Delta _2}{\rm{ }} \buildrel \Delta \over = {M_1}\left[ {{\beta _n}\left( {K_n^2 \!+\! 4{K_n} + 2} \right) \!+\! {{\left( {{K_n} \!+\! 1} \right)}^2}\sum\limits_{ i \ne n}^N {{\beta _i}} } \right].
\end{align}
\end{lemma}

\begin{IEEEproof}
When $M$ grows large, we employ \cite[Lemma 1]{zhang2014power} and approximate (\ref{test6}) as
\begin{equation}\label{test8}
R_{{\rm{P,n}}}^{{\rm{MRC}}}  \approx   {{{\log }_2}\left( {1 + \frac{{{p_u}\mathbb{E}\left\{{{\left| {\boldsymbol{\rm g}_{n0}^H{\boldsymbol{\rm g}_{n0}} + \alpha \boldsymbol{ \rm g}_{n1}^H{\boldsymbol{\rm g}_{n1}}} \right|}^2}\right\}}}{{{\mathbb{E}\left\{\Psi _1\right\}}}}} \right)}  .
\end{equation}
With the help of \cite[Lemma 3]{zhang2014power}, the expectations in (\ref{test8}) can be derived as
\begin{align}
\mathbb{E} \left\{ {{{\left\| {{{\mathbf{g}}_{nj}}} \right\|}^2}} \right\} &= {\beta _n}{M_j},\label{eq:expectation2}\\
\mathbb{E} \left\{ {{{\left\| {{{\mathbf{g}}_{nj}}} \right\|}^4}} \right\} &= \beta _n^2\left( {M_j^2 + \frac{{2{M_j}{K_n} + {M_j}}}{{{{\left( {{K_n} + 1} \right)}^2}}}} \right),\label{eq:expectation1}\\
\mathbb{E}\left\{ {{{\left| {{\mathbf{g}}_{nj}^H{{\mathbf{g}}_{ij}}} \right|}^2}} \right\} &\!=\! \beta _n^2\left( {\frac{{{K_n}{K_i}\phi _{ni,j}^2 \!+\! {M_j}\left( {{K_n} \!+\! {K_i}} \right) \!+\! {M_j}}}{{\left( {{K_n} \!+\! 1} \right)\left( {{K_i} \!+\! 1} \right)}}} \right)\notag.
\end{align}
Moreover, considering \eqref{eq:inter_noise}, the last term in the denominator of \eqref{test8} is given by \cite[Eq. (15)]{zhang2016spectral}
\begin{align}\label{eq:expectation_quantization}
\mathbb{E}\left\{ {{\mathbf{g}}_{n1}^H{\text{diag}}\left( {{{\mathbf{G}}_1}{\mathbf{G}}_1^H} \right){{\mathbf{g}}_{n1}}} \right\} &= \beta _n^2{M_1}\frac{{K_n^2 + 4{K_n} + 2}}{{{{\left( {{K_n} + 1} \right)}^2}}} \notag \\
&+ {\beta _n}{M_1}\sum\limits_{i \ne n}^N {{\beta _i}}.
\end{align}
Substituting \eqref{eq:expectation1} to \eqref{eq:expectation_quantization} into \eqref{test8}, the proof can be concluded after some simplifications.
\end{IEEEproof}

For the case of perfect CSI at the BS with $M \rightarrow \infty$, the impact of the Rician $K$-factor $K_n$, transmit power $p_u$, large-scale fading $\beta_n$, and the proportion of mixed-resolution ADCs $\kappa$ on the achievable rate $R_{{\rm{P,n}}}^{{\rm{MRC}}}$ can be revealed in Lemma \ref{lemma1}. It is clear from (\ref{test7}) that the achievable rate increases by using more BS antennas. We further show the relationship between the achievable rate and key parameters by providing the following important analysis.


\newtheorem{remark}{Remark}
\begin{remark}\label{Remark1}
The derived result in Lemma \ref{lemma1} is general. It is interesting to note that if all users have infinite Rician $K$-factors, the approximate uplink achievable rate \eqref{test7} approaches a constant value as
\begin{align}\label{K_inf}
&R_{{\rm{P,n}}}^{{\rm{MRC}}} \to  {\log _2}\Bigg( 1 \notag\\
& \!+\! \frac{{{p_u}{\beta _n}{{\left( {{M_0} \!+\! \alpha {M_1}} \right)}^2}}}{{{p_u}\sum\limits_{i \ne n}^N {{\beta _i}\left( {{\phi _{ni,0}^2} \!+\! {\alpha ^2}{\phi _{ni,1}^2}} \right)}  \!+\! {M_0} \!+\! \alpha {M_1} \!+\! \alpha \rho {p_u}{M_1}\sum\limits_{i = 1}^N {{\beta _i}} }} \Bigg),
\end{align}
which reveals that there is an achievable rate limit in strong LoS scenarios. When $\alpha=1$, \eqref{K_inf} reduces to the special case of unquantized massive MIMO systems \cite[Eq. (44)]{zhang2014power}.
In addition, \eqref{test7} includes the achievable rate of Rayleigh fading channels as a special case. More specifically, with ${K_n} = {K_i} = 0$, \eqref{test7} can be reduced to
\begin{align}\label{K_0}
R_{{\rm{P,n}}}^{{\rm{MRC}}} \to {\log _2}\left( {1 + M\frac{{\left( {\alpha  + \rho \kappa } \right){\beta _n}}}{{\frac{1}{{{p_u}}} + \sum\limits_{i = 1,i \ne n}^N {{\beta _i}}  + \frac{{2\alpha \rho \left( {1 - \kappa } \right)}}{{\alpha  + \rho \kappa }}{\beta _n}}}} \right),
\end{align}
which is consistent with \cite[Eq. (9)]{tan2016spectral}.
\end{remark}
\begin{remark}\label{Remark2}
For the power-scaling law, let us define ${p_u} = {{{E_u}} \mathord{\left/{\vphantom {{{E_u}} M}} \right.\kern-\nulldelimiterspace} M}$ with a fixed $E_u$ and a large $M$, then the limit of (\ref{test7}) converges to
\begin{equation}\label{test9}
R_{{\rm{P,n}}}^{{\rm{MRC}}} \to {\log _2}\left( {1 + {E_u}{\beta _n}\left( {\rho \kappa  + \alpha } \right)} \right),\ \ {as}\ M \to\infty.
\end{equation}
{It is clear from (\ref{test9}) that the Rician $K$-factor is irrelevant to the exact limit of the uplink rate of mixed-ADC massive MIMO systems with MRC receivers due to the channel hardening effect \cite{hochwald2004multiple}.} In contrast to ideal ADCs, the proportion of the high-resolution ADCs $\kappa$ and the distortion factor of the low-resolution ADC $\rho$ have effects on the limit rate when scaling down the transmit power proportionally to $1/M$. More specifically, the achievable rate can be improved by increasing $\kappa$. Adopting a simple transformation, $\rho \kappa  + \alpha  = \left( {1 - \kappa } \right)\alpha  + \kappa $, we can clearly see that the achievable rate is a monotonic increasing function of $\alpha$. This means that we can increase the achievable rate by using higher quantization bits in the $2M_1$ low-resolution ADCs.
\end{remark}

\subsection{Imperfect CSI}
In practical massive MIMO systems, it is generally difficult to obtain accurate CSI at the BS. Therefore, for the most practical and general case of imperfect CSI, we consider a transmission within the coherence interval $T$ and use $\tau$ symbols for pilots. {The CSI for each antenna can be obtained by using the high-resolution ADCs in a round-robin manner, which has been clearly explained in \cite{liang2016mixed}.} The power of pilot symbols is ${p_p} \buildrel \Delta \over = \tau {p_u}$, while the minimum mean-squared error (MMSE) estimate of $\boldsymbol{\rm G}$ is given by $\boldsymbol{\rm {\hat G}}$. Let $\boldsymbol{\Xi}  = \boldsymbol{\rm {\hat G}} - \boldsymbol{\rm G}$ be the channel estimation error matrix, which is independent of $\boldsymbol{\rm {\hat G}}$ due to the properties of MMSE estimation \cite{ngo2013energy}. The elements of $\boldsymbol{\Xi}$ has a variance of $\sigma _n^2 \buildrel \Delta \over = \frac{{{\beta _n}}}{{\left( {1 + {p_p}{\beta _n}} \right)\left( {{K_n} + 1} \right)}}$  \cite[Eq. (21)]{zhang2014power}.
Following a similar analysis in \cite{ngo2013energy}, the achievable rate of the $n$th user under imperfect CSI is given by
\begin{equation}\label{proofeq1}
R_{{\rm{IP,n}}}^{{\rm{MRC}}} = \mathbb{E}\left\{ {{{\log }_2}\left( {1 + \frac{{{p_u}{{\left| {\boldsymbol{\rm  {\hat g}}_{n0}^H{\boldsymbol{\rm {\hat g}}_{n0}} + \alpha \boldsymbol{ \rm  {\hat g}}_{n1}^H{\boldsymbol{\rm  {\hat g}}_{n1}}} \right|}^2}}}{{{\Psi _2}}}} \right)} \right\},
\end{equation}
where
\begin{align}
{\Psi _2} &={p_u}\sum\limits_{i = 1,i \ne n}^N {{{\left| {{\boldsymbol {\rm{{\hat g}}}}_{n0}^H{\boldsymbol{\rm {\hat g}}_{i0}} + \alpha {\boldsymbol{\rm{{\hat g}}}}_{n1}^H{\boldsymbol{\rm{\hat  g}}_{i1}}} \right|}^2}} +{\left| {{\boldsymbol{\rm{{\hat g}}}}_{n0}^H{{\boldsymbol{\rm{n}}}_0} } \right|^2}+\alpha{\left| {  {\boldsymbol{\rm{{\hat g}}}}_{n1}^H{{\boldsymbol{\rm{n}}}_1}} \right|^2} \notag\\
&+{p_u}\left( {{{\left\| {{{{\boldsymbol{\rm{\hat  g}}}}_{n0}}} \right\|}^2} + {\alpha ^2}{{\left\| {{{{\boldsymbol{\rm{\hat  g}}}}_{n1}}} \right\|}^2}} \right)\sum\limits_{i = 1}^N {\sigma _i^2}\notag\\
&+ \alpha \rho{p_u} \boldsymbol{ \rm {\hat g}}_{n1}^H{\rm{diag}}\left( {\left( {{\boldsymbol {\rm {\hat G}}_1} - {\boldsymbol{\Xi} _1}} \right){\left( {{\boldsymbol {\rm {\hat G}}_1} - {\boldsymbol{\Xi} _1}} \right)^H}} \right){\boldsymbol{\rm {\hat g}}_{n1}}.
\end{align}

\begin{lemma}  \label{lemma2}
For mixed-ADC massive MIMO systems using MRC with imperfect CSI over Rician fading channels, the approximate achievable rate of the $n$th user is given by
\begin{align}\label{test10}
&R_{{\rm{IP,n}}}^{{\rm{MRC}}}\approx {\log _2}\Bigg( 1 + \notag \\
& \frac{{{p_u}{\beta _n}\left[ {{{\left( {{M_0} + \alpha {M_1}} \right)}^2}{{\left( {{K_n} + {\eta _n}} \right)}^2} + \left( {{M_0} + {\alpha ^2}{M_1}} \right)\left( {\eta _n^2 + 2{K_n}{\eta_n}} \right)} \right]}}{{\left( {{M_0} + \alpha {M_1}} \right)\left( {{K_n} + {\eta _n}} \right)\left( {{K_n} + 1} \right)\left( {1 + {p_u}\sum\limits_{i = 1}^N {\sigma _i^2} } \right) + {p_u}\left( {{K_n} + 1} \right)\sum\limits_{ i \ne n}^N {{\Delta _3}}  + \alpha \rho {p_u}{M_1}{\Delta _4}}} \Bigg),
\end{align}
where
\setcounter{equation}{28}
\begin{align}
{\eta _n} &\buildrel \Delta \over = \frac{{{p_p}{\beta _n}}}{{1 + {p_p}{\beta _n}}},\\
{\Delta _3} &\buildrel \Delta \over = \frac{{{\beta _i}}}{{{K_i} + 1}}{K_n}{K_i}\left( {\phi _{ni,0}^2 + {\alpha ^2}\phi _{ni,1}^2} \right) \notag\\
&+ \frac{{{\beta _i}\left( {{M_0} + {\alpha ^2}{M_1}} \right)}}{{{K_i} + 1}}\left( {{K_i}{\eta _n} + {K_n}{\eta _i} + {\eta _n}{\eta _i}} \right),\\
{\Delta _4} &\buildrel \Delta \over = {\beta _n}\left( {K_n^2 + 4{K_n}{\eta _n} + 2\eta _n^2} \right) \notag\\
& + \left( {{K_n} + 1} \right)\left( {{K_n} + {\eta _n}} \right)\sum\limits_{i \ne n}^N {\frac{{{\beta _i}\left( {{K_i} + {\eta _i}} \right)}}{{{K_i} + 1}}} .
\end{align}
\end{lemma}

\begin{IEEEproof}
When the number of BS antennas $M$ grows without bound, we employ \cite[Lemma 1]{zhang2014power} to approximate (\ref{proofeq1}) as
\begin{equation}\label{test11}
R_{{\rm{IP,n}}}^{{\rm{MRC}}}  \approx   {{{\log }_2}\left( {1 + \frac{{{p_u}\mathbb{E}\left\{{{\left| {\boldsymbol{\hat {\rm g}}_{n0}^H{\boldsymbol{\hat{\rm g}}_{n0}} + \alpha \boldsymbol{\hat{ \rm g}}_{n1}^H{\boldsymbol{\hat{\rm g}}_{n1}}} \right|}^2}\right\}}}{{{\mathbb{E}\left\{\Psi _2\right\}}}}} \right)}  .
\end{equation}
Utilizing \cite[Lemma 5]{zhang2014power}, the expectations in (\ref{test11}) can be found as
\begin{align}
\mathbb{E} \left\{ {{{\left\| {{{\mathbf{\hat {g}}}_{nj}}} \right\|}^2}} \right\} &= {\beta _n}\left( {\frac{{{M_j}{K_n}}}{{{K_n} + 1}} + \frac{{{M_j}{\eta _n}}}{{{K_n} + 1}}} \right)\label{eq:expectation4}\\
\mathbb{E} \left\{ {{{\left\| {{{\mathbf{\hat {g}}}_{nj}}} \right\|}^4}} \right\} &= \frac{{\beta _n^2}}{{{{\left( {{K_n} + 1} \right)}^2}}}\bigg[ M_j^2K_n^2 + \left( {M_j^2 + {M_j}} \right)\eta _n^2\notag \\
&+ \left( {2{M_j}{K_n} + 2M_j^2{K_n}} \right){\eta _n} \bigg]\label{eq:expectation5}\\
\mathbb{E}\left\{ {{{\left| {{\mathbf{\hat {g}}}_{nj}^H{{\mathbf{\hat {g}}}_{ij}}} \right|}^2}} \right\} &= \frac{{{\beta _n}{\beta _i}}}{{\left( {{K_n} + 1} \right)\left( {{K_i} + 1} \right)}}\bigg[ {K_n}{K_i}\phi _{ni,j}^2 \notag \\
&+ {M_j}{K_i}{\eta _n} + {M_j}{K_n}{\eta _i} + {M_j}{\eta _n}{\eta _i} \bigg]\label{eq:expectation6}.
\end{align}
Furthermore, with the help of the known variance and mean of the elements of $\boldsymbol{\Xi}$, and \cite[Eq. (22)]{zhang2016spectral}, we can easily derive the expectation of the last term in the denominator of \eqref{proofeq1} as
\begin{align}\label{eq:expectation_quantization2}
&\mathbb{E}\left\{\boldsymbol{ \rm {\hat g}}_{n1}^H{\rm{diag}}\left( {\left( {{\boldsymbol {\rm {\hat G}}_1} - {\boldsymbol{\Xi} _1}} \right){\left( {{\boldsymbol {\rm {\hat G}}_1} - {\boldsymbol{\Xi} _1}} \right)^H}} \right){\boldsymbol{\rm {\hat g}}_{n1}} \right\} \notag\\
&=\mathbb{E}\left\{\boldsymbol{ \rm {\hat g}}_{n1}^H{\rm{diag}}\left( {{\boldsymbol {\rm {\hat G}}_1}{{\boldsymbol {\rm {\hat G}}_1}^H}} \right){\boldsymbol{\rm {\hat g}}_{n1}} \right\} + \mathbb{E}\left\{\boldsymbol{ \rm {\hat g}}_{n1}^H{\rm{diag}}\left( {{\boldsymbol{\Xi} _1}{ {\boldsymbol{\Xi} _1}^H}} \right){\boldsymbol{\rm {\hat g}}_{n1}} \right\} \notag\\
&=   \frac{{{M_1}\beta _n^2}}{{{{\left( {{K_n} + 1} \right)}^2}}}\left( {K_n^2 + 4{K_n}{\eta _n} + 2\eta _n^2} \right) \notag\\
&+ \frac{{{M_1}{\beta _n}\left( {{K_n} + {\eta _n}} \right)}}{{{K_n} + 1}}\sum\limits_{i = 1,i \ne n}^N {\frac{{{\beta _i}\left( {{K_i} + {\eta _i}} \right)}}{{{K_i} + 1}}}   \notag\\
&  +  {\frac{{{M_1}{\beta _n}\left( {{K_n} + {\eta _n}} \right)}}{{{K_n} + 1}}\sum\limits_{i = 1}^N {\sigma _i^2} } .
\end{align}
Substituting \eqref{eq:expectation4} to \eqref{eq:expectation_quantization2} into \eqref{proofeq1}, the proof can be concluded after some simplifications.
\end{IEEEproof}

Note that Lemma \ref{lemma2} presents a closed-form approximation for the achievable rate of mixed-ADC massive MIMO systems with imperfect CSI over Rician fading channels. {The derived approximate rate expressions are more and more accurate as the number of antennas increases and the Rician $K$-factor decreases. This finding coincides with the one presented in \cite[Figs. 2-3]{zhang2014power}}. It is clear that the channel estimation error has a baneful effect on the achievable rate. We now perform further analysis based on the uplink rate approximation \eqref{test10}.

\newcounter{mytempeqncnt1}
\begin{figure*}[!b]
\normalsize
\setcounter{mytempeqncnt1}{\value{equation}}
\setcounter{equation}{36}
\hrulefill
\begin{align}\label{Rayleigh_IP}
R_{{\rm{IP,n}}}^{{\rm{MRC}}} \approx {\log _2}\left( {1 + \frac{{{p_u}{\beta _n}{\eta _n}\left[ {{{\left( {{M_0} + \alpha {M_1}} \right)}^2} + \left( {{M_0} + {\alpha ^2}{M_1}} \right)} \right]}}{{\left( {{M_0} + \alpha {M_1}} \right)\left( {1 + {p_u}\sum\limits_{i = 1}^N {\tilde \sigma _i^2} } \right) + {p_u}\sum\limits_{i \ne n}^N {{{\tilde \Delta }_3}}  + \alpha \rho {p_u}{M_1}{{\tilde \Delta }_4}}}} \right),
\end{align}
\setcounter{equation}{\value{mytempeqncnt1}}
\vspace*{2pt}
\end{figure*}

\begin{remark}  \label{Remark3}
For unquantized massive MIMO systems with ideal ADCs ($\alpha  = 1$) over Rician fading channels, \eqref{test10} can reduce to \cite[Eq. (82)]{zhang2014power}. For the special case of Rayleigh fading channels, \eqref{test10} reduces to
\begin{equation}\label{Rayleigh_IP}
R_{{\rm{IP,n}}}^{{\rm{MRC}}} \approx {\log _2}\left( {1 + \frac{{{p_u}{\beta _n}{\eta _n}\left[ {{{\left( {{M_0} + \alpha {M_1}} \right)}^2} + \left( {{M_0} + {\alpha ^2}{M_1}} \right)} \right]}}{{\left( {{M_0} + \alpha {M_1}} \right)\left( {1 + {p_u}\sum\limits_{i = 1}^N {\tilde \sigma _i^2} } \right) + {p_u}\sum\limits_{i \ne n}^N {{{\tilde \Delta }_3}}  + \alpha \rho {p_u}{M_1}{{\tilde \Delta }_4}}}} \right),
\end{equation}
where $\tilde \sigma _i^2 = \frac{{{\beta _n}}}{{1 + {p_p}{\beta _n}}}$, ${{\tilde \Delta }_3} = {\beta _i}{\eta _i}\left( {{M_0} + {\alpha ^2}{M_1}} \right)$, and ${{\tilde \Delta }_4} = {\beta _n}{\eta _n} + \sum\limits_{i = 1}^N {{\beta _i}{\eta _i}} $. To the best of our knowledge, \eqref{Rayleigh_IP} is new and has not been derived before.
\end{remark}

\begin{remark}  \label{Remark4}
With ${p_u} = {{{E_u}} \mathord{\left/{\vphantom {{{E_u}} {{M^\gamma }}}} \right.\kern-\nulldelimiterspace} {{M^\gamma }}}$ and $M \to \infty $, the limit of \eqref{test10} converges to
\setcounter{equation}{37}
\begin{align}\label{test12}
R_{{\rm{IP,n}}}^{{\rm{MRC}}} &\to {\log _2}\left( {1 + \left( {\rho \kappa  + \alpha } \right)\frac{{{E_u}{\beta _n}\left( {{M^\gamma }{K_n} + \tau {E_u}{\beta _n}} \right)}}{{{M^{2\gamma  - 1}}\left( {{K_n} + 1} \right)}}} \right),\notag \\
&\;\;\;  \text{as}\ M \to\infty.
\end{align}
It is worth pointing out that the power scaling law with imperfect CSI depends on the value of $n$th user's Rician $K$-factor $K_n$. For Rayleigh fading channels ($K_n  = 0$), the user transmit power $p_u$ can be at most cut down by a factor of $1/\sqrt{M}$. If there is a LoS component ($\kappa  > 0$), the transmit power of each user can be scaled down proportionally to $1/{M}$.

For the special case of low-resolution ADC architectures ($\kappa = 0$), \eqref{test12} reduces to
\begin{align}\label{eq:pure_R_IP}
R_{{\rm{IP,n}}}^{{\rm{MRC}}} &\to {\log _2}\left( {1 + \frac{{\alpha {E_u}{\beta _n}\left( {{M^\gamma }{K_n} + \tau {E_u}{\beta _n}} \right)}}{{{M^{2\gamma  - 1}}\left( {{K_n} + 1} \right)}}} \right), \notag \\
&\;\;\;  \text{as}\ M \to\infty.
\end{align}
which is consistent with \cite[Eq. (23)]{zhang2016spectral}. For the unquantized massive MIMO systems ($\kappa = 1, \alpha = 1$), \eqref{test12} converges to a constant value as
\begin{align}
R_{{\rm{IP,n}}}^{{\rm{MRC}}} &\to {\log _2}\left( {1 + \frac{{{E_u}{\beta _n}\left( {{M^\gamma }{K_n} + \tau {E_u}{\beta _n}} \right)}}{{{M^{2\gamma  - 1}}\left( {{K_n} + 1} \right)}}} \right),\notag \\
& \;\;\;  \text{as}\ M \to\infty,
\end{align}
which is the exact achievable rate of unquantized massive MIMO systems as derived in \cite{zhang2014power}.
\end{remark}

\begin{remark}\label{Remark5}
In contrast to the case of perfect CSI, $R_{{\rm{IP,n}}}^{{\rm{MRC}}}$ is affected by the Rician $K$-factor $K_n$. When $\gamma  = 1$, \eqref{test12} reduces to a constant value as
\begin{equation}\label{test13}
R_{{\rm{IP,n}}}^{{\rm{MRC}}} \to {\log _2}\left( {1 + \frac{{\left( {\rho \kappa  + \alpha } \right){E_u}{\beta _n}{K_n}}}{{{K_n} + 1}}} \right),\;\;\;  \text{as}\ M \to\infty.
\end{equation}
Note that \eqref{test13} converges to \eqref{test9} for very strong LoS scenarios. i.e., ${K_n} \to \infty $, where the channel estimation becomes far more robust. This is because items in the fading matrix that were random before becomes deterministic now. Also, similar to the perfect CSI case \eqref{test9}, \eqref{test13} clearly indicates that as either the number of high-resolution ADCs ($\kappa$) or the quantization bit of low-resolution ADCs ($\alpha$) gets larger, the achievable rate can be improved monotonously.
\end{remark}

\section{Numerical Results}\label{se:numerical_results}
In this section, we provide simulated and analytical results on the uplink sum achievable rate of mixed-ADC massive MIMO systems, which is defined as $R = \sum\nolimits_{n = 1}^N {{R_n}} $. The energy efficiency is also examined in detail. First, we assume the users are uniformly distributed in a hexagonal cell with a radius of 1000 meters, while the minimum distance between the user to the BS is ${r_{\min }} = 100$ meters. Furthermore, the pathloss is expressed as $r_n^{ - v }$ with ${r_n}$ denoting the distance between the $n$th user to the BS and $v=3.8$ be the pathloss exponent, respectively. The shadowing is modeled as a log-normal random variable ${s_n}$ with standard deviation ${\delta _s}=8 $ dB. Therefore, the large-scale fading is given by ${\beta _n} = {s_n}{\left( {{{{r_n}} \mathord{\left/ {\vphantom {{{r_n}} {{r_{\min }}}}} \right. \kern-\nulldelimiterspace} {{r_{\min }}}}} \right)^{ - v}}$. Finally, ${\theta _n}$ are assumed to be uniformly distributed within the interval $\left[ {{{ - \pi } \mathord{\left/
 {\vphantom {{ - \pi } {2,{\pi  \mathord{\left/
 {\vphantom {\pi  2}} \right.
 \kern-\nulldelimiterspace} 2}}}} \right.
 \kern-\nulldelimiterspace} {2,{\pi  \mathord{\left/
 {\vphantom {\pi  2}} \right.
 \kern-\nulldelimiterspace} 2}}}} \right]$.

\begin{figure}[!t]
\centering
\includegraphics[scale=1]{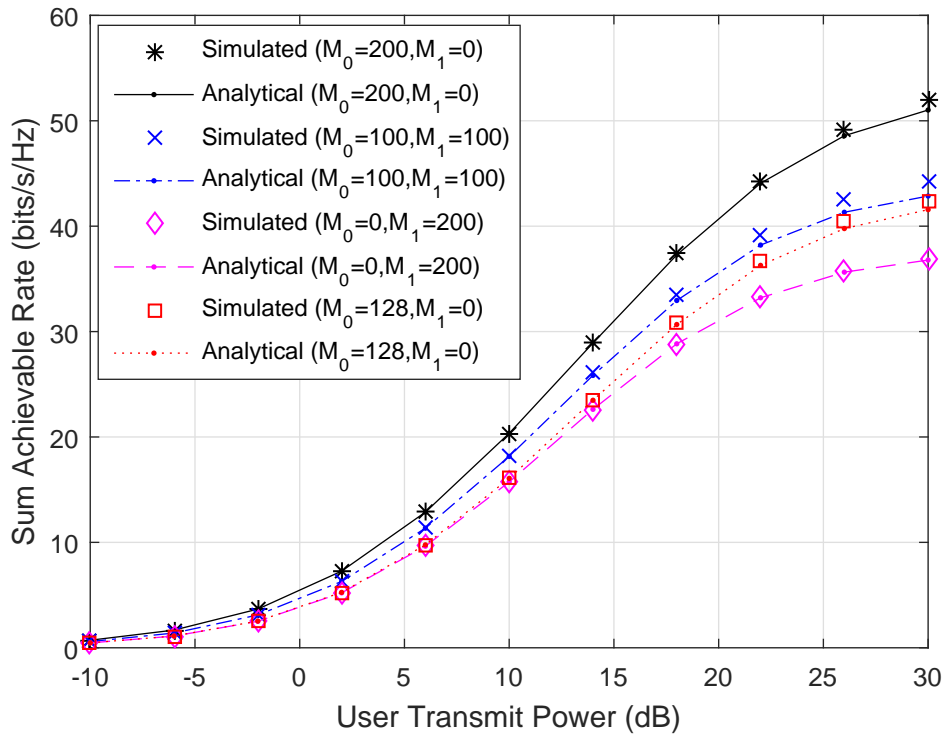}
\caption{Sum achievable rate of mixed-ADC massive MIMO systems over Rician fading channels against the user transmit power for $K = 10 $ dB, $N = 10$ and $b=1$.}
\label{SE_pu}
\end{figure}
\subsection{Achievable rate}
In Fig. \ref{SE_pu}, the simulated and the analytical achievable rate (\ref{test7}) of mixed-ADC massive MIMO systems with perfect CSI are plotted against the user transmit power $p_u$. Herein, we consider the following four cases: 1)${M_0} = 200,\,{M_1} = 0$; 2)${M_0} = 100,\,{M_1} = 100$; 3)${M_0} = 0,\,{M_1} = 200$; 4)${M_0} = 128,\,{M_1} = 0$. Fig. \ref{SE_pu} shows that the analytical and simulated curves are very tight in all considered cases, which confirms the correctness of our derived results. Moreover, the sum achievable rates in four cases are similar at low SNR. While as $p_u$ increases, the gap among different cases becomes relatively large. It is clear that the massive MIMO systems with ideal ADCs (Case 1) gains superior achievable rate performance compared to the one with mixed-ADCs (Case 2) and the one with low-resolution ADCs (Case 3). With similar hardware cost as Case 2, the achievable rate of Case 4 is lower due to a smaller number of BS antennas are used.


\begin{figure}[!t]
\centering
\includegraphics[scale=1]{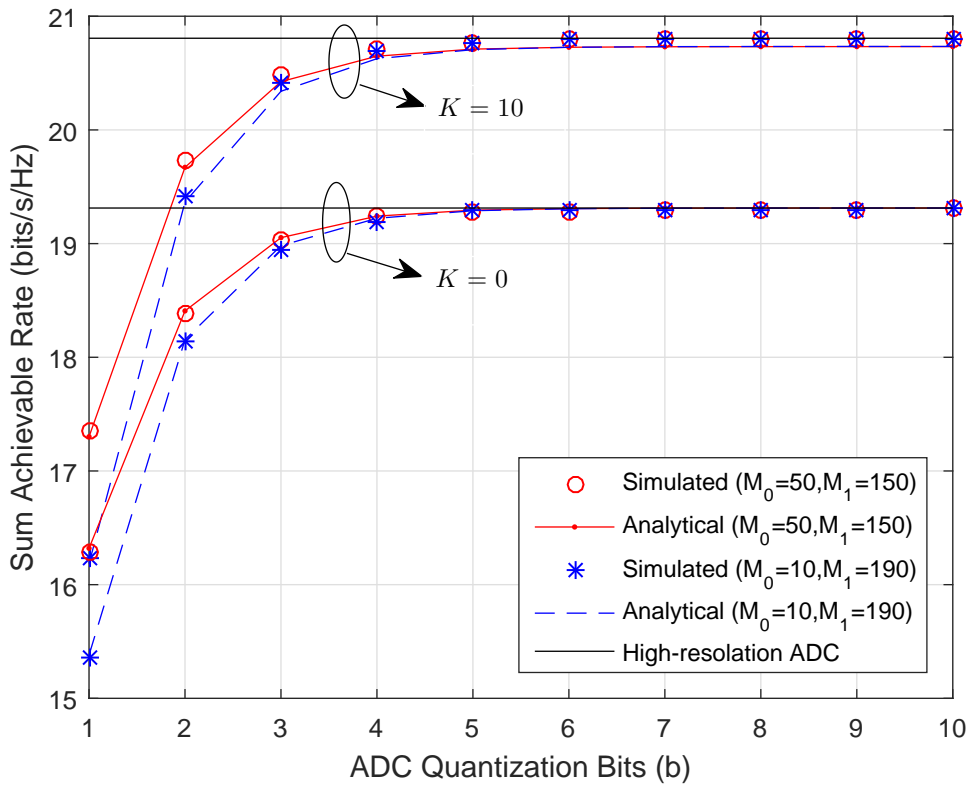}
\caption{Sum achievable rate of mixed-ADC massive MIMO systems over Rician fading channels against the ADC quantization bits for $N = 10$, and ${p_u} = 10 $ dB.}
\label{SE_b_K}
\end{figure}

Fig. \ref{SE_b_K} presents the simulated and approximate sum achievable rate of mixed-ADC massive MIMO systems for various numbers of ADC quantization bits. Without loss of generality, we assume that all users have the same Rician $K$-factor as $K_n = K$. The simulation results notably coincide with the derived results for all ADC quantization bits. We also find from Fig. \ref{SE_b_K} that higher sum rate can be achieved by employing more high-resolution ADCs. Furthermore, the achievable rate increases with the number of quantization bits ($b$), and converges to a limited achievable rate with high-resolution ADCs. For Rayleigh fading channels ($K=0$), the mixed-ADC architecture can achieve the same sum achievable rate by employing 5 bits, while for Rician fading channels ($K=10$), more ADC quantization bits are needed, which is in agreement with Remark \ref{Remark1}. This is true because the quantization noise is larger with more received power in LoS dominating scenarios.

\begin{figure}[!t]
\centering
\includegraphics[scale=1]{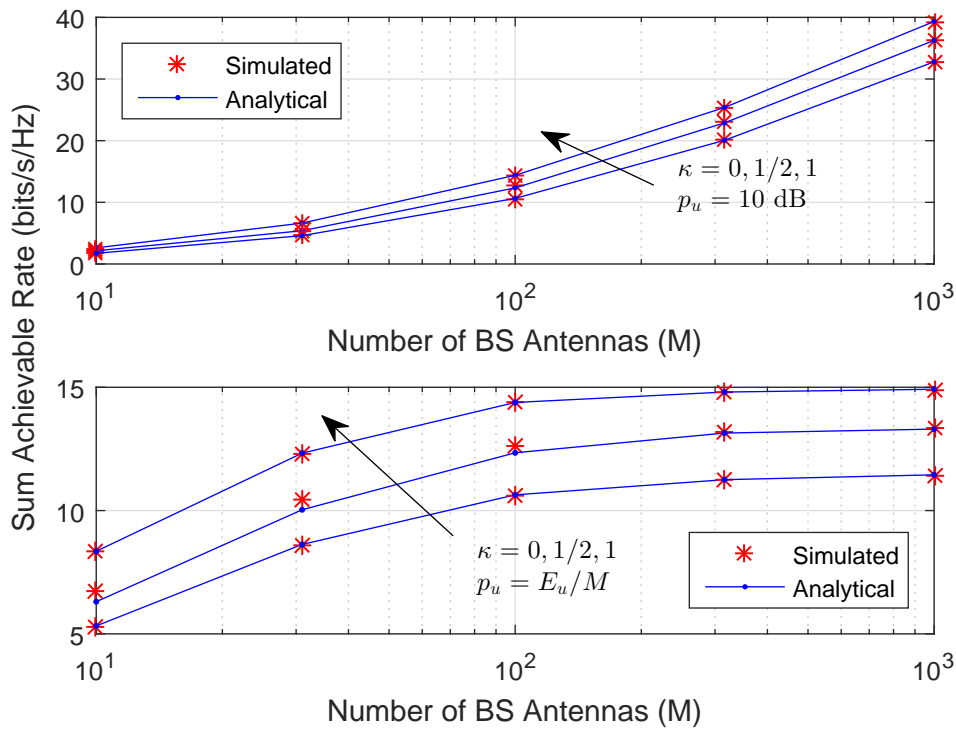}
\caption{Sum achievable rate of mixed-ADC massive MIMO systems over Rician fading channels against the number of BS antennas for $K = 10{\rm{dB}}$, $N = 10$, and $b=1$.}
\label{SE_M}
\end{figure}

In Fig. \ref{SE_M}, we investigate the power scaling law of mixed-ADC massive MIMO systems over Rician fading channels with perfect CSI. The proportions of the number of high-resolution ADCs in the mixed-ADC architecture are considered as $\kappa  = 0$, ${1 \mathord{\left/ {\vphantom {1 2}} \right. \kern-\nulldelimiterspace} 2}$, and 1, respectively. As predicted in Remark \ref{Remark2}, a higher proportion of the number of high-resolution ADCs improves the achievable rate. For fixed values of ${p_u=10}$ dB, all curves grow without bound by increasing $M$; while for variable values of ${p_u}=E_u / M$, the corresponding curves eventually saturate with an increased $M$. These results validate that we can scale down the transmitted power of each user as $E_u / M$ for the perfect CSI case over Rician fading channels.

\begin{figure}[tbp]
\centering
\includegraphics[scale=1]{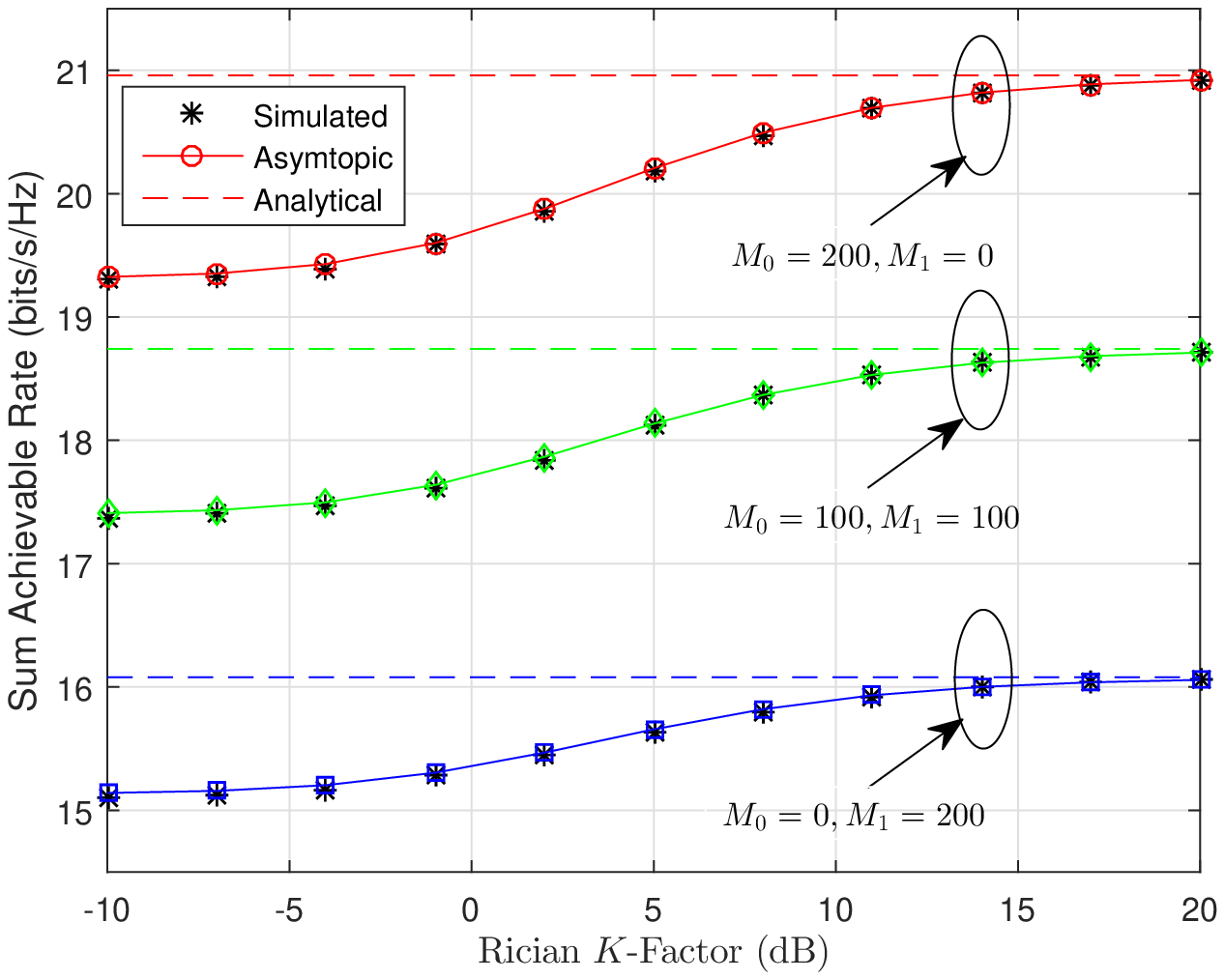}
\caption{Sum achievable rate of mixed-ADC massive MIMO systems over Rician fading channels against the Rician $K$-factor for $p_u= 10 $ dB, $N = 10$, and $b=1$.}
\label{SE_K}
\end{figure}

A beneficial sum achievable rate improvement arising with a higher Rician $K$-factor is illustrated in Fig. \ref{SE_K}. Here, the following three cases considered in Fig. \ref{SE_pu} are investigated: 1)${M_0} = 200,\,{M_1} = 0$; 2)${M_0} = 100,\,{M_1} = 100$; 3)${M_0} = 0,\,{M_1} = 200$. The three constant curves are obtained by the analytical expression \eqref{K_inf}. As expected, the sum rate approach a fixed value as $K \rightarrow \infty$. In addition, placing more high-resolution ADCs at the BS greatly enhances the overall performance of the sum achievable rate.

\begin{figure}[tbp]
\centering
\includegraphics[scale=1]{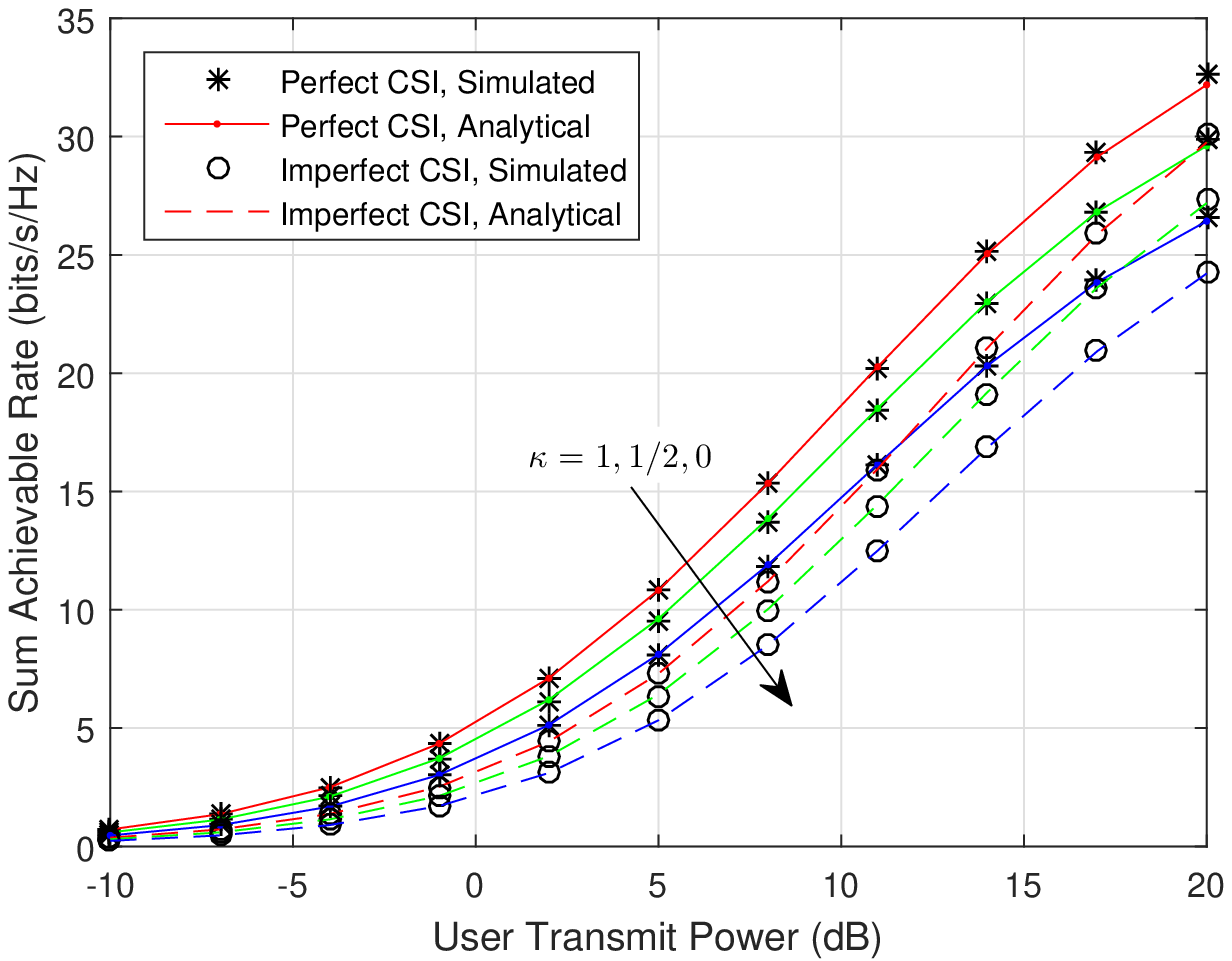}
\caption{Sum achievable rate of mixed-ADC massive MIMO systems over Rician fading channels against the user transmit power for $K = 0{\rm{dB}}$, $M= 200$, $N = 10$, and $b=1$.}
\label{SE_pu_kappa}
\end{figure}

Fig. \ref{SE_pu_kappa} presents the sum achievable rate of different ADC architectures with both perfect and imperfect CSI over Rician fading channels. As can be seen, the sum achievable rate grows obviously without bound as the user transmit power $p_u$ increases for both perfect and imperfect CSI. An insightful observation is that the sum achievable rates of all cases are similar at low SNR. However, the gaps become large when the SNR increases.

\subsection{Energy Efficiency}
{Until now, we have investigated the achievable rate of mixed-ADC massive MIMO systems over Rician fading channels. As expected, the sum rate of unquantized massive MIMO systems outperforms the one with mixed-ADCs, at the cost of higher hardware cost and power consumption. Therefore, there should be a fundamental trade-off between the energy efficiency and power consumption for practical massive MIMO systems. Due to this reason, the energy efficiency of mixed-ADC massive MIMO systems over Rician fading channels is studied in this section.}

{Let us first consider a generic power consumption model of the mixed-ADC architecture. The energy efficiency ${\eta _\text{EE}}$ is defined as \cite{mo2016hybrid}
\begin{equation}
{\eta _\text{EE}} = \frac{{RW}}{{{P_\text{total}}}}\;\;\;\text{{bit} / {Joule}},
\end{equation}
where {$R$ denotes the sum rate}, $W$ denotes the transmission bandwidth assumed to be 1 GHz and $P_\text{total}$ is the total consumed power, which can be expressed as \cite{mendez2016hybrid}
\begin{align}\label{eq:power_model}
{P_\text{total}} &= {P_\text{LO}} + M\left( {{P_\text{LNA}} + {P_\text{H}} + 2{P_\text{M}}} \right) \notag\\
&+ 2{M_0}\left( {{P_\text{AGC}} + P_\text{ADC}^H} \right) + 2{M_1}\left( {c{P_\text{AGC}} + P_\text{ADC}^L} \right) + P_\text{BB},
\end{align}
where $P_\text{LO}$, $P_\text{LNA}$, $P_\text{H}$, $P_\text{M}$, $P_\text{AGC}$, $P_\text{ADC}^H$, $P_\text{ADC}^L$, $P_\text{BB}$ denote the power consumption in the LO, LNA, $\pi/2$ hybrid and LO buffer, Mixer, AGC, high-resolution ADCs, low-resolution ADCs, and baseband processor, respectively. In addition, $c$ is the flag related to the quantization bit of low-resolution ADCs and is given by
\begin{align}\label{eq:c}
c = \left\{ {\begin{array}{*{20}{c}}
  {0,\;\;\;b = 1}, \\
  {1,\;\;\;b > 1}.
\end{array}} \right.
\end{align}
Furthermore, the power consumed in the ADCs can be expressed in terms of the number of quantization bits as
\begin{equation}
{P_\text{ADC}} = {\text{FOM}_W} \cdot {f_s} \cdot {2^b},
\end{equation}
where $f_s$ is the Nyquist sampling rate, $b$ is the number of quantization bits, and $\text{FOM}_W$ is Walden's figure-of-merit for evaluating the power efficiency with ADC's resolution and speed \cite{walden1999analog}. In the simulation, the consumed power in the RF components are assumed to be ${P_\text{LO}} = 22.5{\rm{mW}}$ \cite{scheir200852}, ${P_\text{LNA}} = 5.4{\rm{mW}}$ \cite{shang20128mw}, ${P_\text{H}} = 3{\rm{mW}}$ \cite{marcu2011generation}, ${P_\text{AGC}} = 2{\rm{mW}}$ \cite{shang20128mw}, ${P_\text{M}} = 0.3{\rm{mW}}$ \cite{jin20117db}, ${P_\text{BB}} = 200{\rm{mW}}$ \cite{mendez2016hybrid}. With state-of-the-art technology, the power consumption value of $\text{FOM}_W$ at 1 GHz bandwidth is $\text{FOM}_W = 5\sim15$ fJ/conversion-step \cite{WinNT}.}

\begin{figure}[t]
\centering
\includegraphics[scale=1]{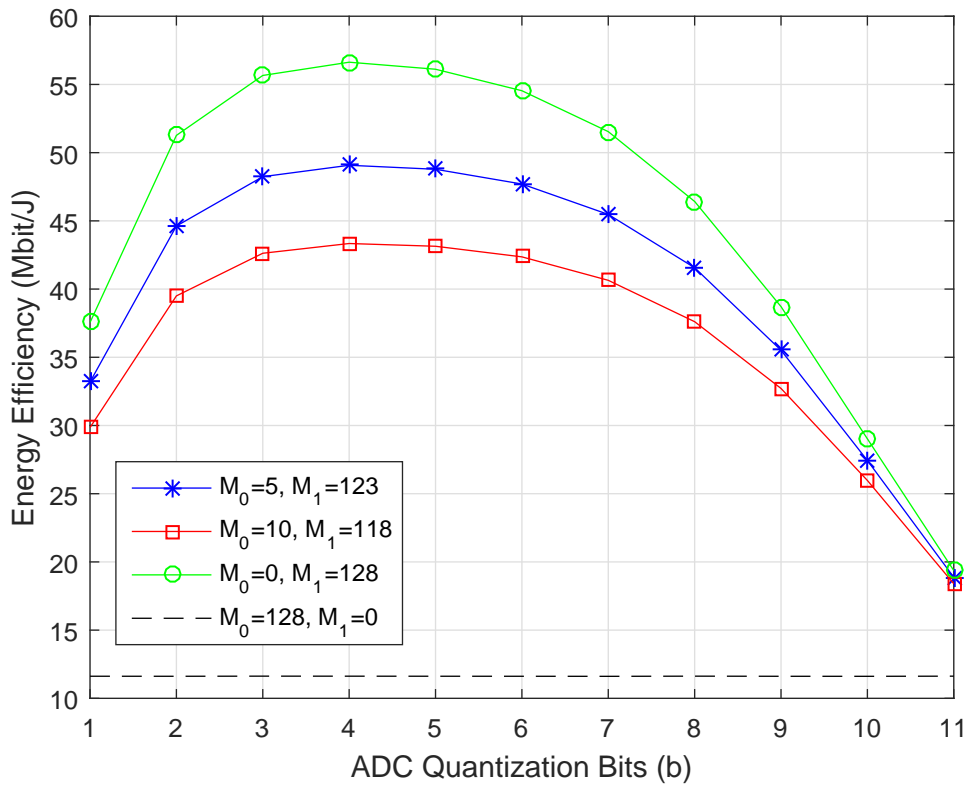}
\caption{Energy efficiency of mixed-ADC massive MIMO systems over Rician fading channels against different ADC quantization bits for $K = 10 $ dB, $N = 10$, and ${p_u} = -10 $ dB.}
\label{EE_b}
\end{figure}

\begin{figure}[tbp]
\centering
\includegraphics[scale=1]{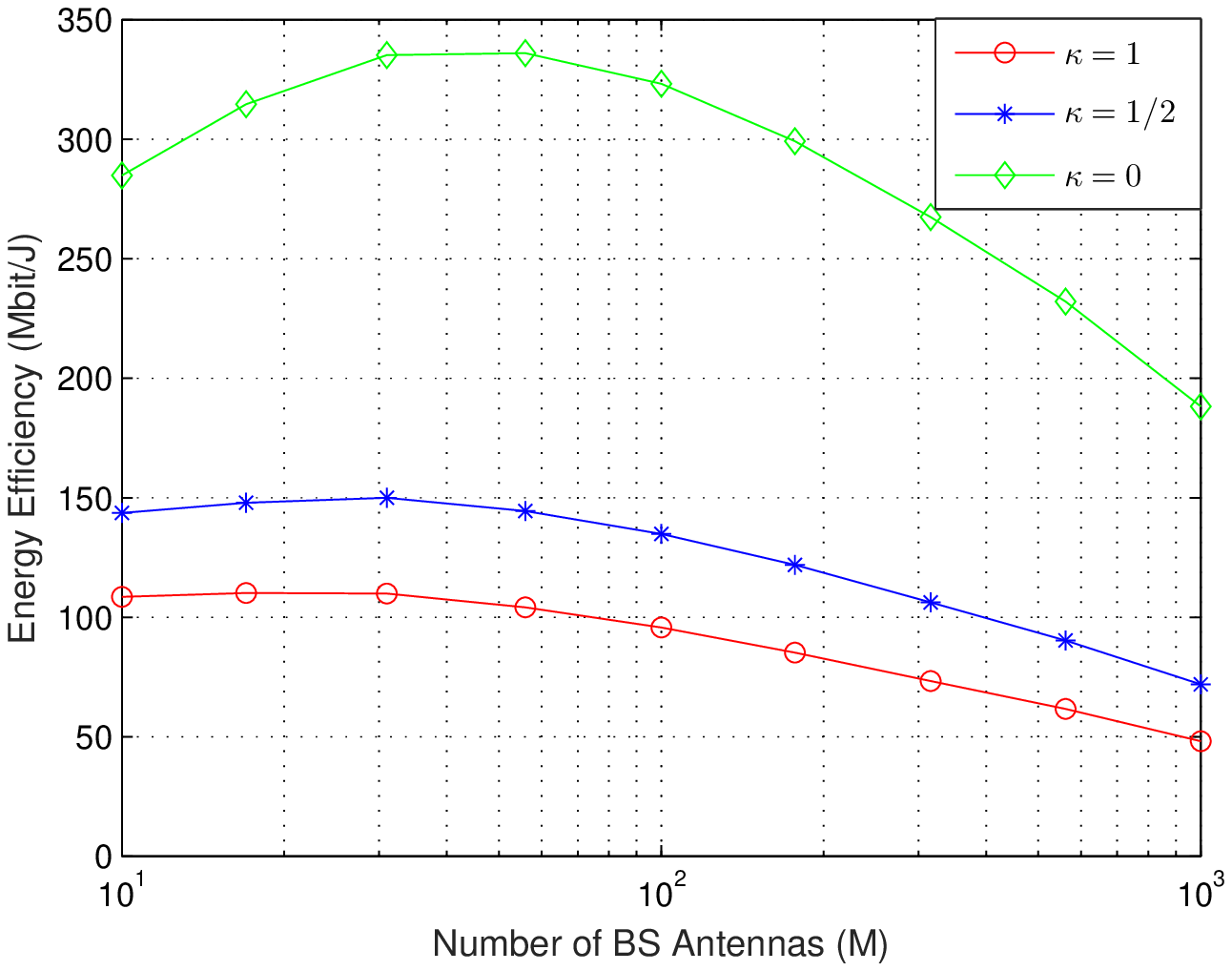}
\caption{Energy efficiency of mixed-ADC massive MIMO systems over Rician fading channels against different numbers of antennas for $K = 10 $ dB, $N = 10$, and ${p_u} = 0 $ dB.}
\label{EE_M}
\end{figure}

\begin{figure}[tbp]
\centering
\includegraphics[scale=1]{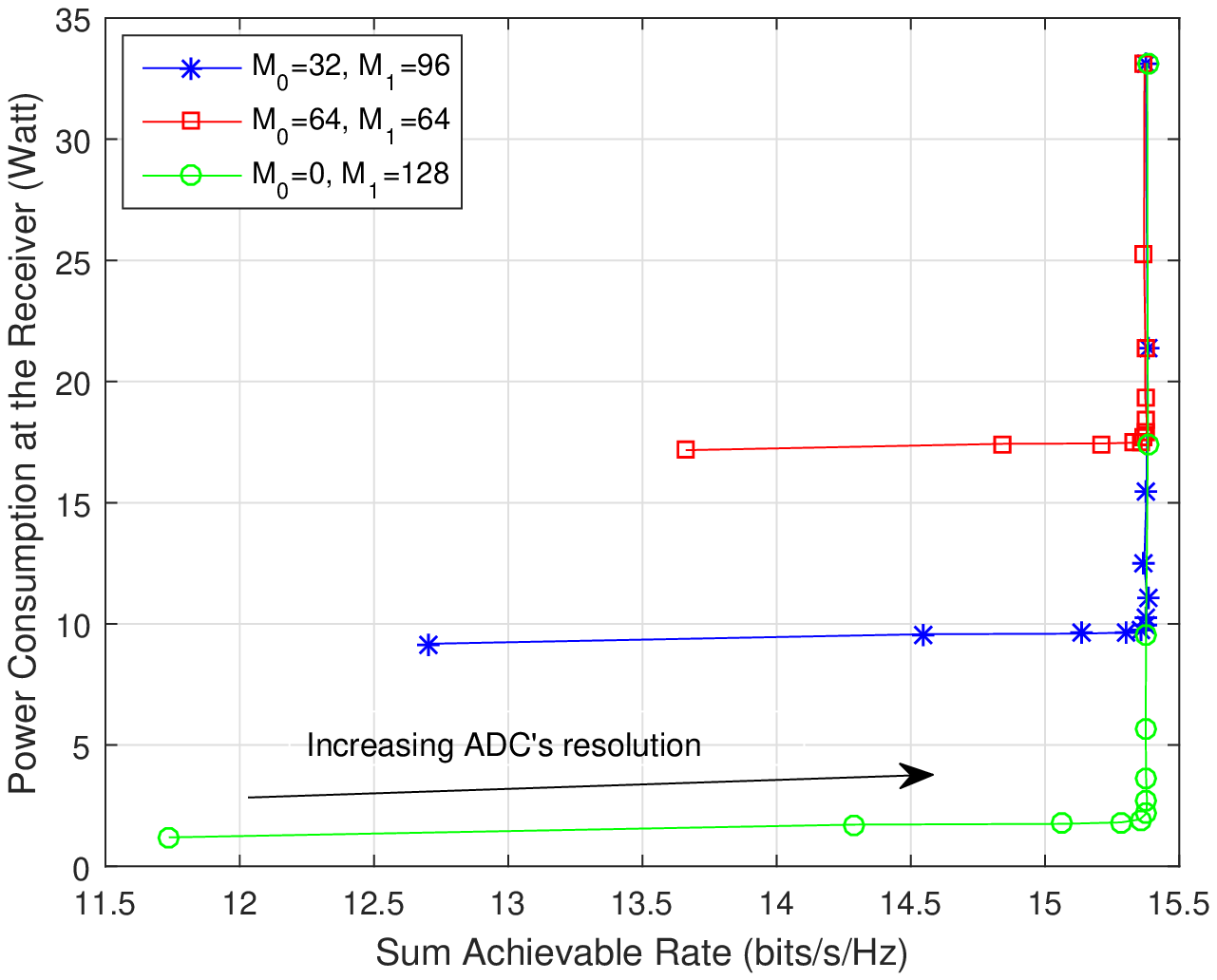}
\caption{Trade-off between power consumption and achievable rate of mixed-ADC massive MIMO systems over Rician fading channels for $K = 10 $ dB, $N = 10$, and ${p_u} = 10 $ dB.}
\label{P_SE_b}
\end{figure}

\begin{figure}[t]
\centering
\includegraphics[scale=1]{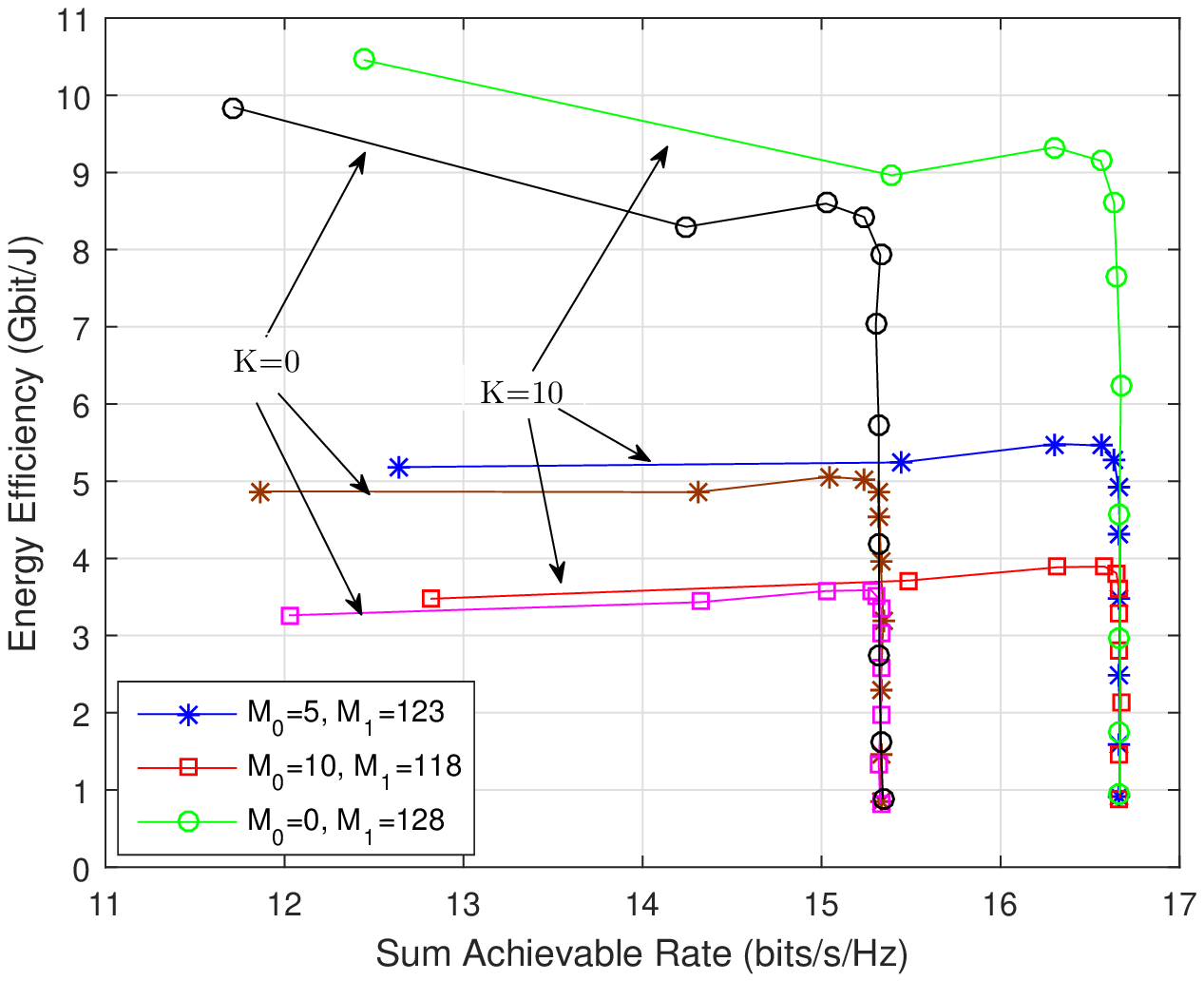}
\caption{Trade-off between energy efficiency and achievable rate of mixed-ADC massive MIMO systems over Rician fading channels for $N = 10$ and ${p_u} = 10 $ dB.}
\label{EE_SE_K}
\end{figure}

{With the power consumption model \eqref{eq:power_model}, the energy efficiency of massive MIMO systems with different ADC architectures and perfect CSI is illustrated in Figs. \ref{EE_b}-\ref{EE_SE_K}.} We find that the energy efficiency of the mixed-ADC architecture can be significantly improved compared with ideal high-resolution ADCs. Moreover, the receiver adopting only low-resolution ADCs has the highest energy efficiency among all architectures. This implies that the energy efficiency decreases with the proportion of the number of high-resolution ADCs ($\kappa $) in the mixed-ADC structure. {Although the low-resolution ADC architecture can achieve better energy efficiency than the mixed-ADC architecture, the spectral efficiency of the low-resolution ADC architecture is much lower than that of the mixed-ADC architecture. Moreover, the channel estimation in the mixed-ADC architecture are more tractable than that in the low-resolution ADC architecture due to the use of partial high-resolution ADCs.} With all results, it is clear that we can achieve a better achievable rate meanwhile obtain a relatively high energy efficiency by using the mixed-ADC architecture.

In Fig. \ref{EE_M}, we analyze the effect of the number of receive antennas on the energy efficiency. Due to the fact that each antenna is connected with one RF chain, it may be wasteful to use more antennas at the BS. The energy efficiency of low-resolution ADCs is largest among others. Indeed, if the BS has only around 30 antennas, the mixed-ADC architecture achieves its optimal energy efficiency for the selecting system parameters.

\begin{figure}[t]
\centering
\includegraphics[scale=1]{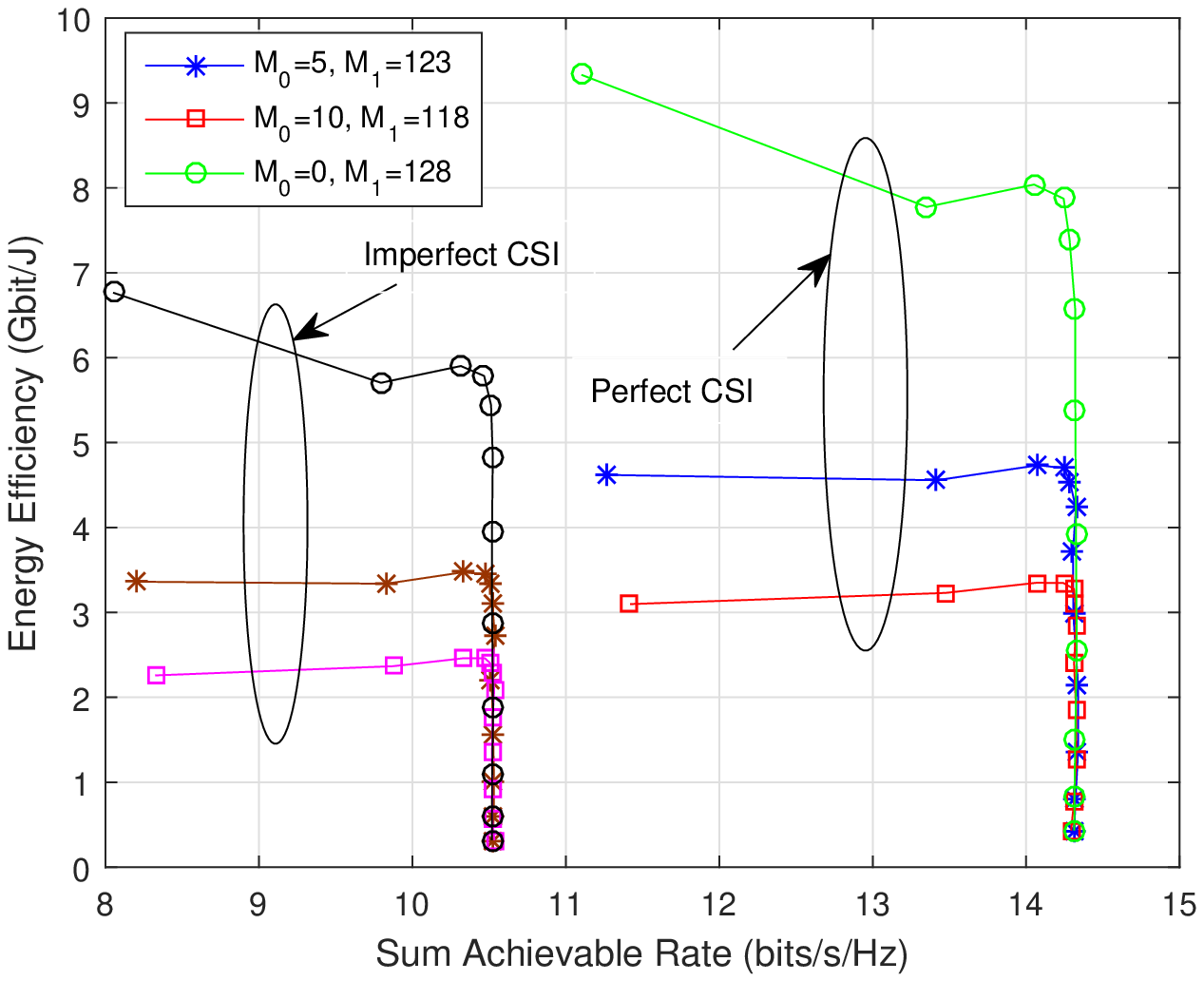}
\caption{Trade-off between energy efficiency and achievable rate of mixed-ADC massive MIMO systems over Rician fading channels for imperfect and perfect CSI, $K = 10 $ dB, $N = 10$ and ${p_u} = 0 $ dB.}
\label{EE_SE_CSI_0dB}
\end{figure}

\begin{figure}[htbp]
\centering
\includegraphics[scale=1]{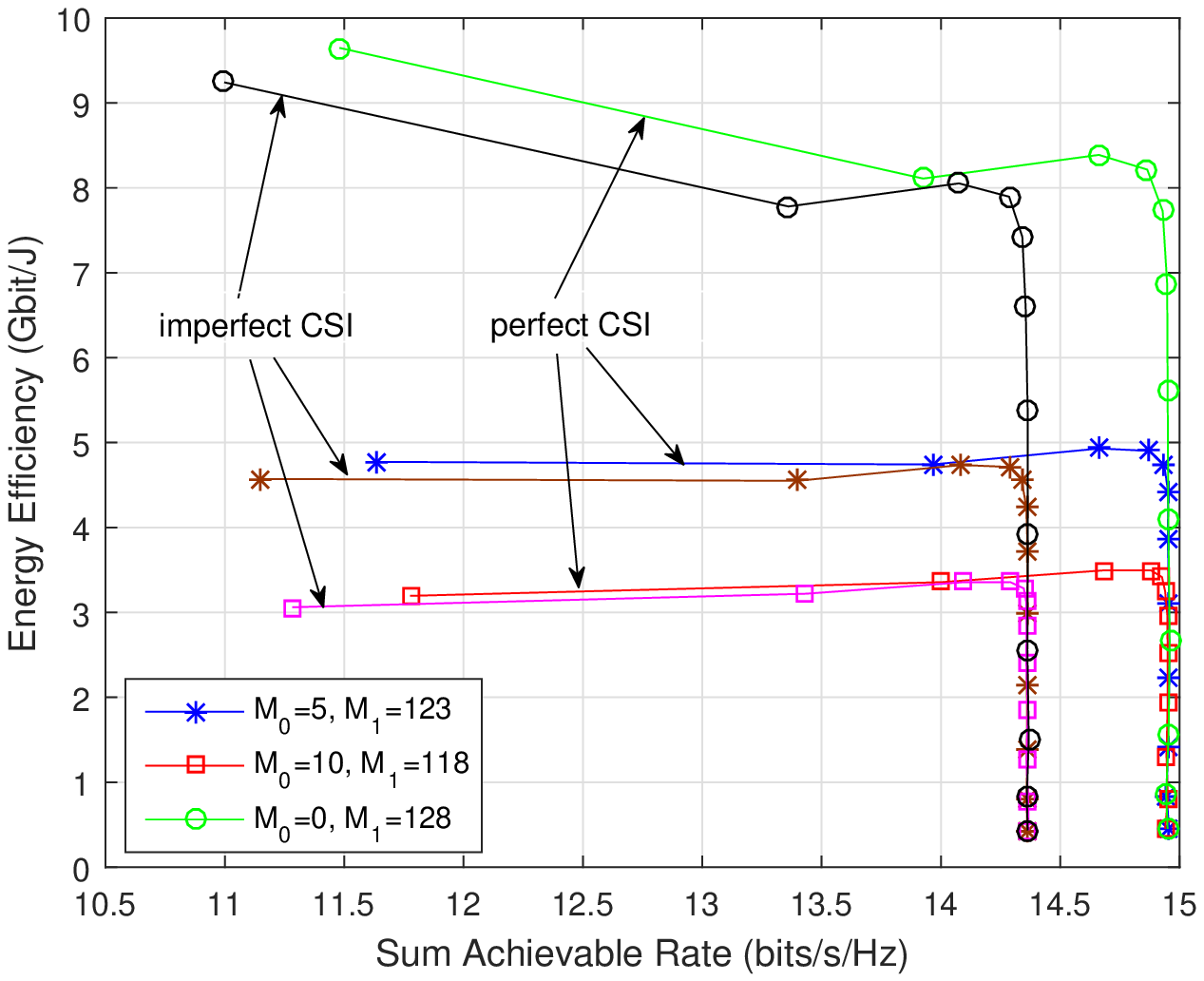}
\caption{Trade-off between energy efficiency and achievable rate of mixed-ADC massive MIMO systems over Rician fading channels for imperfect and perfect CSI, $K = 10 $ dB, $N = 10$ and ${p_u} = 10 $ dB.}
\label{EE_SE_CSI_10dB}
\end{figure}

Next, we present some numerical results in {Figs. \ref{EE_SE_K}-\ref{EE_SE_CSI_10dB}} to illustrate the trade-off between energy efficiency and achievable rate of the mixed-ADC architecture for different portions of high-resolution ADCs ($\kappa$), Rician $K$-factor, and quantization bits of low-resolution ADCs ($b$). In addition, for each figure, we plot curves representing the trade-off performance as we increase the number of ADC quantization bits from 1 to 12, at the increment of 1. Note that the largest value of the achievable rate is given at the rightmost points, while the highest points in the figure correspond to the largest value of the energy efficiency. Therefore, it is better to reach a point to the top-right corner for the corresponding architecture. Moreover, if the area contained at the left of the curve, which can be interpreted as an ``operating region", is bigger, the corresponding architecture is more versatile as a function of the ADC quantization bit.

The first insight obtained from Fig. \ref{P_SE_b} is that the mixed-ADC architecture requires more power than the one with pure low-resolution ADCs, but with a relatively big increase in the achievable rate for small quantization bits. In addition, the gap of power consumption decreases with the increasing ADC's resolution. For the mixed-ADC architecture, the achievable rate significantly increases when the quantization bit increases from 1 to 4, with a negligible increase in the receiver power consumption. However, the power consumption increases significantly when the quantization bit going from 4 to 12 bits. This means that the mixed-ADC architecture can achieve a better trade-off between the power consumption and achievable rate by using 4 bits.

Fig. \ref{EE_SE_K} shows the trade-off between energy efficiency and achievable rate of mixed-ADC architectures for different values of the Rician $K$-factor. We also plot the curves of the pure low-resolution ADC architecture as the evaluation baselines. Note that the signal processing in the pure low-resolution ADC architecture is very complex and time-consuming compared with the mixed-ADC architecture. It is clear from Fig. \ref{EE_SE_K} that the optimal ADC quantization bit is influenced by both the portion of high-resolution ADCs and the Rician $K$-factor. For the mixed-ADC architecture, the energy efficiency only increases up to a certain number of quantization bits, while it decreases at higher ADC quantization bits. This is due to the fact that the achievable rate is a sub-linear increasing function of the quantization bits, whereas the power consumption of ADCs increases exponentially with $b$. As expected, the mixed-ADC architecture can achieve larger operating region when operating over Rician fading channels with stronger LoS component (large values of $K_n$) due to higher achievable rate shown in Fig. \ref{SE_K}.

The effect of imperfect CSI and user transmit power on the trade-off between energy efficiency and achievable rate of mixed-ADC massive MIMO systems has been investigated in Figs. \ref{EE_SE_CSI_0dB} and \ref{EE_SE_CSI_10dB}. We observe that the operating region of imperfect CSI is smaller than the one of perfect CSI. However, this gap is reduced by improving the user transmit power $p_u$ form 0 dB to 10 dB. This means that more user transmit power is needed to combat the effect of channel estimation error.

\section{Conclusions}\label{se:conclusion}
In this paper, the performance of mixed-ADC massive MIMO systems over Rician fading channels is investigated. We derive novel and closed-form approximate expressions for the achievable rate for large-antenna limit. The cases of both perfect and imperfect CSI are incorporated in our analysis. With similar hardware cost, the mixed-ADC architecture can achieve larger sum rate than the ideal-ADC architecture. For mixed-ADC massive MIMO systems over Rician fading channels, the user transmit power can be at most cut down by a factor of $1/M$. In contrast to the case of perfect CSI, the power scaling law in the imperfect CSI case is related to the Rician $K$-factor. For both perfect and imperfect CSI, the achievable rate limit converges to a constant as $K \to \infty $. In addition, the mixed-ADC architecture can have a large operating region with few high-resolution ADCs. More energy efficiency can be achieved when operating over stronger LoS scenarios. It is worth mentioning that the derived results are general and can include many previously reported ones as special cases. Finally, we conclude that practical massive MIMO can achieve a considerable performance with small power consumption by adopting the mixed-ADC architecture for 5G.

\bibliographystyle{IEEEtran}
\bibliography{IEEEabrv,Mixed_ADC}

\end{document}